\documentclass[usenatbib]{mn2e}
\usepackage{graphicx}
\usepackage{amssymb}
\usepackage{lscape}
\usepackage{ulem}
\usepackage{txfonts}
\usepackage{url}

\def\kms{km~s$^{-1}$}

\def\ha{H$\alpha$}
\def\hb{H$\beta$}

\def\SII{S{\sc ii}}
\def\delsfr{$\Delta$ log(SFR)}
\newcommand{\nii}{[N{\sc ii}]}
\newcommand{\oii}{[O{\sc ii}]}
\newcommand{\oiii}{[O{\sc iii}]}

\title[Star formation and metallicities in groups]{The dependence of galaxy group star formation rates and metallicities on large scale environment.}

\author[Scudder et al.] {Jillian M. Scudder$^1$\thanks{jscudder@uvic.ca}, Sara L. Ellison$^1$,  J. Trevor Mendel$^1$\\
$^1$ Department of Physics and Astronomy, University of Victoria, 
Victoria, British Columbia, V8P 1A1, Canada.\\}

\begin{document}

\maketitle

\begin{abstract}
We construct a sample of 75,863 star forming galaxies with robust
metallicity and star formation rate measurements from the Sloan
Digital Sky Survey Data Release 7 (SDSS DR7), from which we select a
clean sample of compact group (CG) galaxies.  The CGs are defined to
be close configurations of at least 4 galaxies that are otherwise
apparently isolated.  Our selection results in a sample of 112
spectroscopically identified compact group galaxies, which can be
further divided into groups that are either embedded within a larger
structure, such as a cluster or large group, or truly isolated
systems.  The compact groups then serve as a probe into the influence
of large scale environment on a galaxy's evolution, while keeping the
local density fixed at high values.  We find that the star formation rates (SFRs) of star forming galaxies in
compact groups are significantly different between isolated and embedded systems.  Galaxies in isolated systems show significantly enhanced SFR, relative to a control sample matched in
mass and redshift, a trend not seen in the embedded systems.  
Galaxies in isolated systems exhibit a median SFR enhancement at fixed stellar
mass of $+0.07\pm0.03$ dex.  
These dependences on large scale environment are small in magnitude relative to the apparent influence of local scale effects found in previous studies, but the significance of the difference in SFRs between our two samples constrains the effect of large scale environment to be non-zero.
We find no significant change in
the gas-phase interstellar metallicity for either the isolated or
embedded compact group sample relative to their controls.  However, simulated samples that include artificial
offsets indicate that we are only sensitive to metallicity changes of
log O/H $>$0.13 dex (at 99\% confidence), which is considerably larger
than the typical metallicity differences seen in previous
environmental studies.

\end{abstract}

\begin{keywords}
galaxies: interactions, galaxies: groups: general, galaxies: evolution, galaxies: abundances
\end{keywords}

\section{Introduction}

The evolution of a galaxy is fundamentally shaped by its environment.
Qualitative evidence of the influence of environment has been known
since the 1930s \citep[e.g.,][]{Hubble1931}; the large surveys now
available have added their statistical weight to its influence,
allowing a quantification of environmental changes.  Nearly every
observable quantity of a galaxy has been shown to vary with
environment, including star formation rates (SFRs), morphologies,
colours, active galactic nucleus (AGN) fractions, and mean stellar
mass \citep[e.g.,][]{Dressler1980, Balogh2004, Kauffmann2004,
Croton2005, Baldry2006, Park2007}.  
The primary mechanisms that can operate in denser environments
 are strangulation, harassment, and ram-pressure stripping \citep[e.g.,][]{Gunn1972,
Larson1980, Moore1996}.  However, disentangling which of these might
be a fundamental variable and which are corollary effects is a
non-trivial task \citep[e.g.,][]{Whitmore1991, Blanton2005,
Blanton2007, Skibba2009}.  Moreover, relatively simple processes such
as individual galaxy--galaxy interactions become complicated by the
influence of a local, high density environment.  For example, at low
densities, close galaxy pairs show enhanced SFRs \citep{Lambas2003,
Alonso2004, Ellison2010}.  However, once the galaxy pair is found within a high
density environment, the strength of the SFR enhancement diminishes or
becomes undetectable \citep{Alonso2004, Alonso2006,
Baldry2006,Ellison2010}.  Clearly, galaxies are sensitive to their
environment beyond the scale of their nearest neighbour.  The question
remains, how far beyond the nearest neighbour does environment matter?  Is
it merely a local density effect, or does cluster membership, with
scales far beyond the local, also impact a galaxy's evolution?  This
work will attempt to address this question.

Several density trends are well documented in the literature for low redshift galaxy samples.  The morphology-density and colour-density relations describe the declining fractions of blue, late-type galaxies with increasing density \citep{Dressler1980, Postman1984,Balogh1997, Balogh1998, Martinez2002, Tanaka2004, Baldry2006}.  The overall distribution of star formation rates also correlates with density, with the average galactic SFR declining as density increases \citep{Hashimoto1998, Poggianti2008, Kauffmann2004, Cooper2008}.  Interpretations of the SFR-density relation are complicated by the increasing fraction of low- (or non-) star forming elliptical galaxies as density increases.  The contribution of early type galaxies will bias the average SFR to lower values at higher densities \citep[e.g.,][]{Balogh2004}.

The dependence of SFR on density becomes much less clear for only the star forming subsample of galaxies, with many conflicting interpretations.  Some studies find that among the star-forming fraction, there is no residual dependence on density \citep{Couch2001, Balogh2004,Tanaka2004,Weinmann2006a, Patiri2006, Park2007, Peng2010, McGee2011, Ideue2011}.  According to these works, the observed decrease in SFR with increasing density is entirely due to the changing fraction of star-forming galaxies with density, and once this dependence is eliminated, the distribution of SFRs is independent of density.  Other studies counter these results, finding that the population change is not enough to account for the entire SFR-density relation, and that a density dependence remains \citep{Balogh1998, Pimbblet2002, Gomez2003, Welikala2008}. 
Moreover, the scale over which the SFR is influenced by its surroundings is widely debated.   Many authors find that it is primarily on the local scale (generally considered to be of order 1 Mpc or less) that the residual SFR-density dependence is found \citep{Hashimoto1998, Carter2001, Lewis2002, Kauffmann2004, Blanton2006, Blanton2007}.  By contrast, some studies indicate that the environment on scales of order several Mpc is of greater importance than that of sub-Mpc scales \citep{Goto2003, Park2009a}. 

A further probe into the scales involved in environmental effects comes from the gas phase metallicity, which, like the SFR, can be modulated by the gas supply \citep{Ellison2008a, Mannucci2010, Lara-Lopez2010, Yates2011}.  In samples of interacting pairs, changes in the gas phase metallicity and SFR are linked, and have been shown to respond to some of the same physical processes \citep{Scudder2012b}.  Theory offers the picture of an interaction triggering large scale gas flow from the outer regions of a galaxy to the centre, and the enhanced densities of gas sparking star formation \citep{Mihos1996, Barnes1996, Barnes2004, Rupke2010a, Montuori2010, Torrey2012}.  In agreement with this theoretical picture, close galaxy pairs show enhanced SFRs \citep{Larson1978, Donzelli1997, Barton2000, Lambas2003, Alonso2004, Alonso2006, Woods2007, Ellison2008, Ellison2010} and diluted metallicities \citep{ Kewley2006a, Ellison2008, Michel-Dansac2008}.  This indicates that gas phase metallicities ought to change, as the SFRs do, with density.  In confirmation of this idea, studies of galaxies in clusters and other dense environments have shown metal enhancement relative to the field \citep{ Mouhcine2007, Cooper2008,Ellison2009}.

In this paper, we employ a new tactic to investigate the effects of large scale environmental dependences.  We select two samples of galaxies with similarly high small scale densities, but with different large scale environments.  To this end, we use the catalogue of 3491 compact group (CG) galaxies in 828 CGs presented in \citet[henceforth M11]{Mendel2011} as a refinement on the sample of \citet[henceforth M09]{McConnachie2009}.  Although the original CG selection criteria were designed to identify groups that are not part of a larger overdensity (\citealt{Hickson1982}; M09), M11 find that the SDSS CG sample can be divided into two distinct populations.  One population appears to be truly isolated, whereas the other population appears embedded within a large scale structure, but isolated from other galaxies within the group. 
M11 show that despite the strong differences in their large scale environments, the galaxies in these `embedded' and `isolated' groups have similar morphological properties.  In this paper, we investigate the spectroscopic properties of the embedded and isolated CG galaxies, focussing on their star formation rates and metallicities.  The photometric catalogue of CG galaxies is therefore cross-matched with a spectroscopic sample of emission line galaxies from the SDSS DR7 \citep{Abazajian2009}.  In Section \ref{sec:sample}, we describe the sample selection, including metallicity and AGN calibrations.  In Section \ref{sec:offsets}, we describe our methods for quantifying differences between the isolated and embedded CG galaxies and a sample of control galaxies.  In Section \ref{sec:discussion}, we discuss the implications of our results, along with a comparison to previous works, and present our conclusions in Section \ref{sec:conclusions}.

We assume $\Omega_M = 0.3$, $\Omega_{\Lambda}=0.7$ and H$_0$ = 70 \kms\ Mpc$^{-1}$.

\section{Sample Selection}
\label{sec:sample}

Our spectroscopic sample is taken from the publicly available MPA-JHU
SDSS DR7 catalogue\footnote{Available at:
\url{http://www.mpa-garching.mpg.de/SDSS/DR7/raw_data.html}} of
927,552 galaxies. This catalogue provides measurements of up to 12 emission lines per galaxy, corrected for stellar absorption lines and Galactic reddening.  
Of these lines, we will eventually require robust flux measurements to be
present in \ha, \hb, \oii$\lambda3727$, \oiii$\lambda\lambda{4959},
5007$, and \nii$\lambda{6584}$ for metallicity calibrations\footnote{Recently, \citet{Groves2011} have found that the \hb~equivalent widths from this catalogue are systematically underestimated by 0.35\AA.  We do not expect this to affect our results, as all our comparisons are relative, and the underestimation found is constant across the entire sample.}.  As some
of these galaxies are not unique objects within the catalogue, we use
the MPA-JHU's duplicate catalogue\footnote{Available at
\url{http://www.mpa-garching.mpg.de/SDSS/DR7/Data/all_matches_dr7.dat}}
to remove duplicate galaxies from the sample.   In this
section, we describe the quality control measures taken in order to
select a reliable sample of star-forming galaxies with measurable metallicities.

\subsection{Quality Control}

In order to ensure that the flux measurements will be robust for
subsequent metallicity calculations, careful checks were made on the
quality of the flux measurements.  As an initial step, all galaxies
with zero and negative flux values were excluded.  Flux values of zero
were determined to be either due to clipped lines or due to the line
being redshifted out of the spectral range of the SDSS.  Negative flux
values, by contrast, seemed to be largely due to poorly subtracted
Balmer absorption.

A redshift cut was imposed on the galaxies to ensure that the spectral
lines needed for the metallicity calibrations will not have been
shifted out of the spectral range or into a region of the spectrum
dominated by sky line residuals.  The bluest wavelength of the
emission lines needed in our analysis, \oii$\lambda{3727}$, dictates
the lower redshift limit. This line has to be redshifted at least to
the lower limit of the SDSS spectra at 3800 \AA\footnote{We note that relaxing the redshift criteria from 0.02 to 0.01 does not increase the sample size of our final CG sample. (See Figure 3 of M11.) }.  At the other end of
the spectrum, significant sky line residuals appear at $\sim8000~\AA$.
We use \ha ~to set the upper redshift, by requiring that \ha ~cannot
be shifted past 8000 \AA.  These wavelength constraints result in a
redshift range of $0.02 \leq z \leq 0.25$.  Additionally, we require
that the catalogue flag {\tt z\_warning} be set to 0; a non-zero value
for this parameter indicates a bad redshift.

An inspection of the distribution of flux values for emission lines
showed that both the flux values and their errors were bimodal with a
prominent secondary peak at flux values $> 10^{-11}$ erg s$^{-1}$cm$^{-2}$.  A
number of the galaxies with the highest flux values were visually
inspected and the emission line fluxes were re-measured manually. The
manually measured emission line fluxes were found to be within the
bounds of the lower flux peak, indicating that the high flux values
were spurious.  The continuum fluxes and errors were also found to be
bimodal.  The secondary peak in the continuum error ($>10^{-9}$
erg s$^{-1}$cm$^{-2}$) pushes the continuum fluxes, and thus the line flux,
to extremely large values.  Imposing a cut around the main
distribution of continuum flux errors eliminated the majority of the
secondary (high value) flux peak.  The remainder of the high fluxes
were found to be associated with high flux errors, and imposing a flux
error threshold eliminated the last remnants of the secondary peak in
the flux values.  

\begin{figure}
\centerline{\rotatebox{0}{\resizebox{9 cm}{!}
{\includegraphics{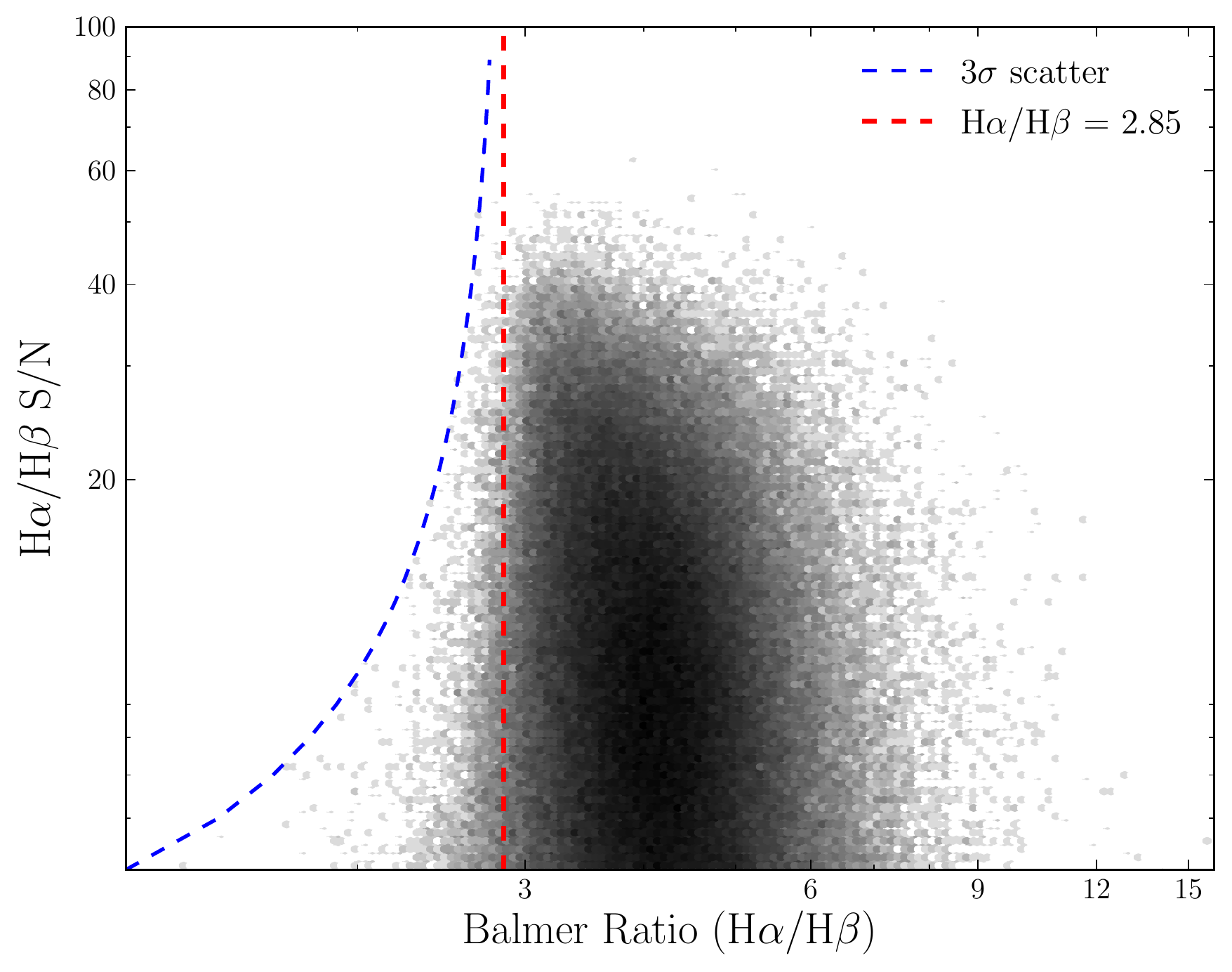}}}}
\caption{ \label{fig:balm} Final sample Balmer ratios vs. Signal to Noise for the
quality controlled sample.  Blue dashed line is the 3$\sigma$ lower limit to the theoretical limit of 2.85 (red vertical line).} 
\end{figure}

It is standard practice to impose a Signal-to-Noise (S/N) cut on
emission line fluxes to ensure high quality metallicity determinations
(e.g.  Kewley \& Ellison 2008).  We adopt a fairly standard S/N cut of
5 on all emission lines needed for our metallicity calibrations.  We
also impose a S/N$>$ 5 cut on the Balmer ratio (the
ratio of \ha\ to \hb\ flux) to minimize the scatter away from the
theoretical Case B recombination limit. Our placement of the S/N cut
in the Balmer ratio is motivated by the data; less stringent S/N cuts
yield a rapid increase in the scatter in the Balmer ratio.  However,
increasing the severity of this cut would not significantly decrease
the scatter, and would serve only to limit our statistics.  As we
expect some random scatter below the theoretical lower limit of 2.85,
we permit 3$\sigma$ scatter below the theoretical limit on the Balmer
ratio (see Figure \ref{fig:balm}).  This ensures that all Balmer
ratios are reasonable, and does not exclude a large number of
galaxies.  A summary of the final quality control cuts is presented in
Table \ref{tab:qualcontrol}.

With robust flux values in hand, we next correct for internal galactic
reddening to obtain intrinsic fluxes, using the Small Magellanic Cloud
(SMC) extinction curve presented in \citet{pei}.  E(B$-$V) values for
the SMC and the Milky Way are very similar at these wavelengths, so
our decision to use the SMC curve does not affect our results.

\begin{table}
\begin{center}
\begin{tabular}{|l|c|}
\hline
\multicolumn{2}{|c|}{Final Quality Control}\\
\hline
Redshift  & $0.02 < z < 0.25$\\
Continuum error & $10^{-19} < C_{err}  < 10^{-15}$\\
Flux error & $10^{-19} <  F_{err}  < 10^{-13}$\\
\ha/\hb~ Signal/Noise &  S/N $> 5$\\
Flux S/N & S/N $>5$\\
\hline
\end{tabular}
\caption{\label{tab:qualcontrol}Final quality control parameters. Continuum and flux errors are in of units erg s$^{-1}$cm$^{-2}$. \ha, \hb, \oii$\lambda3727$, \oiii$\lambda\lambda{4959}, 5007$, and \nii$\lambda{6584}$ must pass these cuts to be considered a reliable detection in that line. }
\end{center}
\end{table}

\subsection{AGN Removal}
\label{sec:agn}

Before metallicities can be calculated, any galaxies containing an Active Galactic Nucleus (AGN)
must be identified and excluded.  AGN are known to alter emission line
strengths, which makes any metallicity calculated for a galaxy hosting
an AGN unreliable. The canonical method of discriminating between
star-forming and AGN galaxies is through the use of emission line
ratios \citep{BPT, vo87}.  In particular, the most broadly used
diagnostic plot is the so-called BPT diagram, after the \citet{BPT} work
in which it was proposed, which plots \oiii$\lambda{5007}$/\hb ~against
\nii$\lambda{6584}$/\ha.  The \oiii ~line is collisionally excited in both
star-forming and AGN regions, so it cannot distinguish between the two
ionisation sources, but rather serves as a metric for the strength of
the ionisation field.  \nii , by contrast, originates in partially
ionised regions, which are found preferentially around AGN sources due
to the power-law radiation they produce.  Photoionisation from O \& B
type stars does not produce a large partially ionised region, so the
\nii ~line strength is relatively small in galaxies dominated by young stellar
light.  The combination of these two line ratios in the BPT diagram
thus allows for differentiation between galaxies dominated by star
formation and those with fluxes dominated by an AGN.  In order for a
galaxy in our sample to pass this diagnostic, it must have reliable
detections in all four of the required emission lines.

There are a number of possible lines drawn on the BPT diagram to
distinguish between star-forming and AGN.  We select the diagnostic of
\citet[K03]{Kauffmann2003}.  K03 is an empirical model, designed
to follow the left-hand wing of the BPT diagram and is defined as follows:
\begin{equation}
\mathrm{log}([\mathrm{OIII}]\lambda{5007} / \mathrm{H}\beta) =\frac{0.61}{\mathrm{log([NII]}\lambda{6584}/\mathrm{H}\alpha) - 0.05} + 1.30.
\label{eq:k03}
\end{equation}
~\\ 

The K03 demarcation follows the star forming wing closely, so the fraction
of AGN galaxies included in the SF sample should be low;
\citet{Stasinska2006} estimates only 3\% contamination from AGN
galaxies. We classify the sample of galaxies as star-forming or AGN
based upon this diagnostic, resulting in a final star forming sample
of 126,756 galaxies.

\subsection{Metallicity Calibrations}
\label{sec:metallicity}

The metallicity calibration adopted here is the adaptation of the
\citet[henceforth KD02]{KD02} recommended method presented in
\citet[KE08]{ke08}, which we refer to as the KD02-KE08 method.  This
calibration does not require the detection of as many strong lines
(and in particular does not require the [\SII] line), which helps to
keep the final sample as large as possible.  The KD02-KE08 calibration has low
intrinsic scatter, and is easily converted into other metallicity
calibrations without significant residuals.

The original KD02 calibration is based on photoionisation models and stellar
population synthesis for high metallicities, and an average of
$R_{23}$ methods for low metallicities, where $R_{23}$ is defined as 

\begin{equation}
R_{23} =\frac{[\mathrm{OII}]\lambda{3727} + [\mathrm{OIII}]\lambda{4959} + [\mathrm{OIII}]\lambda{5007}}{\mathrm{H}_\beta}.
\label{eq:r23}
\end{equation}

The modification implemented by Kewley \& Ellison (2008) is in using 
a different set of $R_{23}$
methods for the low metallicity galaxies.  For high metallicities
(log(\nii/\oii) $> -1.2$), the KD02-KE08 calibration is the same as
presented in KD02.  This calibration requires only \nii ~and \oii
~detections; as long as the galaxy is in the upper branch regime, we
do not require any further strong line detections for the metallicity
calibration.
For log(\nii/\oii) $< -1.2$, the lower branch has changed from the
original average of the \citet{z94} and \citet{m91} calibrations to an
average of the \citet{kk04} and \citet{m91} lower branch calibrations,
all of which are based on the $R_{23}$ and $O_{32}$ diagnostics, where
$O_{32}$ is defined as

\begin{equation}
O_{32} = \frac{[\mathrm{OIII}]\lambda{4959} + [\mathrm{OIII}]\lambda{5007}}{[\mathrm{OII}]\lambda{3727}}.
\label{eq:o32}
\end{equation}

\noindent The
\citet{kk04} calibration takes an iterative approach, solving both for
the ionisation parameter (defined as the number of ionising photons
per number of hydrogen atoms) and the metallicity simultaneously.
As these lower branch calibrations require the \oiii ~lines, galaxies that 
fall in the lower branch have the additional criteria that they must have 
significant detections in \oiii$\lambda\lambda4959,{5007}$.

\subsubsection{Final Metallicity sample}

Our final metallicity sample is 75,863 galaxies and represents both
the sample with which we cross-match the CG sample and the basis of
the pool from which we construct a control sample.  The metallicity sample is
significantly smaller than our initial star forming sample because the
metallicity calculation requires additional emission line detections
beyond those required to classify the galaxy as star forming.  We take
stellar mass\footnote{Unless otherwise stated, all mass values in this
work are stellar masses.}  values as calculated by the MPA/JHU group
from fitting models to the 5-band SDSS DR7 photometry \footnote{A
discussion of their method is available at:
\url{http://www.mpa-garching.mpg.de/SDSS/DR7/mass_comp.html}}.  These
masses are, in general, in good agreement with spectroscopically
derived values calculated by \citet{Kauffmann2003a}.  Star formation
rates within the SDSS fibre (3'') are taken from
\citet{Brinchmann2004}, who use a set of 6 emission line fits to
determine the SFR.  Aperture corrected SFRs are also present in the catalogue.  These SFRs are corrected from fibre to total values
by calculating the galaxy light not contained within the fibre; by
fitting models to the photometry of the galaxy outside the fibre, the
aperture correction can be effectively made (See \citealt{Brinchmann2004} for a full discussion of their aperture correction methodology).  \citet{Salim2007} compare the results of this methodology with the SFRs obtained from UV flux, and find that for star forming galaxies the two methods agree very well, with no bias introduced by the aperture corrections.  We use the aperture-corrected SFR values for the rest of this work.

\subsection{Compact Group Sample}

Our parent CG sample consists of 3491 CG galaxies in 828 CGs
(M11).  This sample is a refinement of the Catalogue A of
M09. Typical intergalactic distances within the
groups are a few tens of kpc, ranging up to 140 kpc. The M09 CG
sample applies a modification of the original criteria laid out by
\citet{Hickson1982} to the SDSS DR6, with isolation, minimum number of
galaxies, and density requirements:

\begin{enumerate}
\item{$N(\Delta m=3) \geq 4$}
\item{$\theta_N \geq 3\theta_G$}
\item{$\mu_e \leq 26.0$ mag arcsec$^{-2}$}
\end{enumerate}

$N(\Delta m=3)$ is the number of galaxies within 3 magnitudes of the
brightest galaxy within the group, $\mu_e$ is the surface brightness
of the group, $\theta_G$ is the angular size of the group, and
$\theta_N$ is the size of the circle beyond which there are no
galaxies within the 3 magnitudes required for group membership.  These
criteria guarantee that there must be at least 4 galaxies in the group
and the group itself must be separated by at least 3 times its own
radius from any other equally bright galaxies.  The surface brightness
criterion guarantees the compactness of the group.  Applying these
criteria, M09 find 2297 CGs, containing 9713 galaxies, down to a
limiting magnitude of $r=18$.  However, there is no redshift criterion
in the identification process for these galaxies; galaxies identified
as a CG are done so based only on their photometry, regardless of
whether or not they have concordant redshifts. Approximately 50\% of the original 9713 galaxies have reliable spectroscopic redshifts in the DR7 (M11). 

In order to address the issue of false (due to projections) CGs in the
M09 sample, M11 apply a statistical likelihood
restriction to the master CG sample. Briefly, by looking at the
probability density functions for the combinations of spectroscopic
and photometric redshifts available for CG galaxies found within a
given group, a large fraction of interloping galaxies can be removed.  If
this process of interloper rejection results in the group failing the
richness criterion used to define the sample, the entire group is
rejected.  M11 estimate that there could be
20\%-30\% contamination remaining in the cleaned sample.
M11 identified a distinction in the distribution of
distances between the CG and the next nearest cluster or rich group.
Approximately 50\% of the cleaned CG sample can be classified as
within $\leq$ 1 Mpc $h^{-1}$ of a rich group, using the
\citet{Tago2010} catalogue compiled from the DR7 (see Figure 5 in
M11). The other half of the sample is distributed at
distances $> 1$ Mpc away from group structures.  Placing a dividing
line in the CGs at the minimum between the two distributions,
M11 describe two populations: `embedded' and
`isolated' CGs.  We adopt this terminology for the galaxies residing
within those groups.  M11 shows that the galaxies
residing within the two density bins have systematically different
photometric properties (i.e., an increase in the fraction of blue
galaxies within isolated CGs).

We cross-match the galaxies 
belonging to either embedded or isolated CGs with our final
metallicity and SFR sample to select only the galaxies residing in CGs that
also have robust metallicities and SFRs.  This cross-matching results in our
final CG galaxy sample of 112 galaxies.  Splitting the sample into
embedded and isolated CGs results in 62 galaxies in isolated CGs, and
50 galaxies in CGs embedded within a large scale structure.  For a list of the embedded and isolated CG galaxies and their properties, see Tables \ref{tab:embedded_CG} \& \ref{tab:isolated_CG} respectively.  SDSS thumbnails of 4 random embedded CGs and 4 isolated CGs are presented in Figures \ref{fig:emb_mosaic} \& \ref{fig:iso_mosaic} respectively.

\subsection{Matching to Controls}

In order to make an effective comparison between our CG sample of
galaxies and a control sample of galaxies, we must eliminate any 
major sources of bias.  
The three major sources of bias are stellar mass, redshift, and environment \citep{Perez2009}.
As we are looking for an effect that varies with environment, we cannot match in that property, but we wish to match a control sample to the CG sample in stellar mass and redshift, such that the distribution of stellar mass and redshift in the control match the distributions in the CG sample.  
Our control pool is defined as any galaxy in the final metallicity sample that is not flagged as belonging to a CG.

Galaxies in our CG sample are matched to galaxies in the control pool
simultaneously in total stellar mass and redshift, in a similar way to
\citet{Ellison2008}.  Briefly, this algorithm finds the best
simultaneous match in mass and redshift to each galaxy in the CG
sample.  Once every galaxy has been matched to a control, a
Kolmogorov-Smirnov (KS) test is used on the total distributions of the CG
sample and control.  If the KS test finds that the distributions are
consistent with being drawn from the same parent distribution at
$>30\%$, then the matching continues without replacement until a maximum of 50
control galaxies have been matched to each CG galaxy in our sample, or
the KS-test results in a probability $<30\%$.  Since the pool of
possible control galaxies is large compared to the CG galaxy sample,
the matching procedure reaches the full capacity of 50 matches per CG
galaxy.  Our final CG samples of 62 isolated CGs and 50 embedded CGs
thus have control samples of 3100 and 2500 galaxies respectively.
The masses and redshifts of the control galaxies are typically
matched to within 0.05 dex and 0.003 respectively of the CG galaxy value.
The E(B$-$V) values of the galaxies are found to be consistent between all four samples when a KS-test is applied.

\begin{figure}
\centerline{\rotatebox{0}{\resizebox{9cm}{!}
{\includegraphics{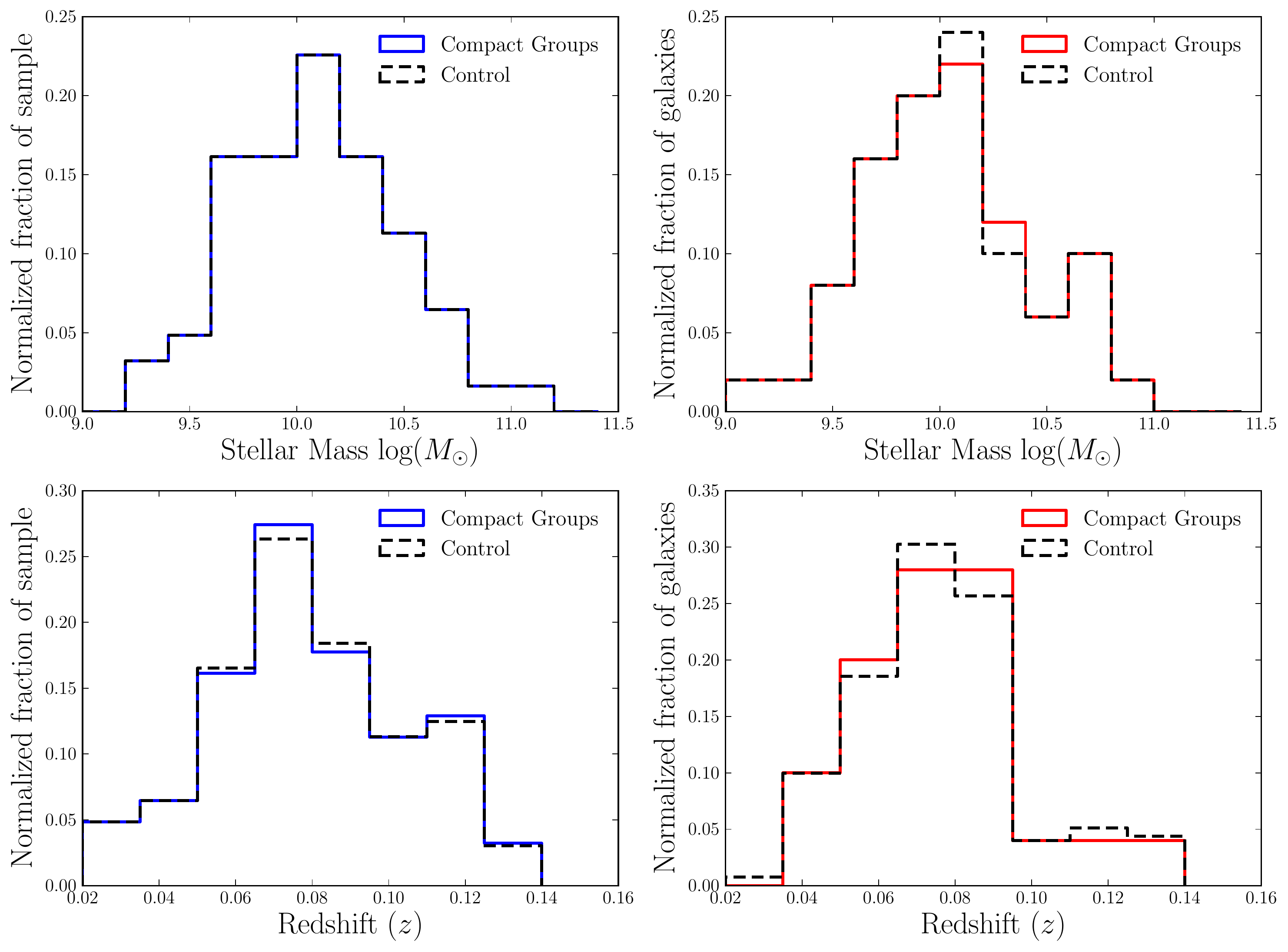}}}}
\caption{\label{fig:match} Normalized distributions of isolated (left panels) and embedded (right panels) CG galaxies and their controls for total stellar mass (upper panels) and redshift (lower panels).  Galaxies are matched to controls in total stellar mass and redshift simultaneously until 50 control galaxies are matched to each group galaxy.  KS-test probabilities that the samples are drawn from the same population are $>99\%$ for the control and CG galaxy distributions in all panels.}
\end{figure}

The resulting normalised distributions of mass and redshift are shown in Figure
\ref{fig:match}.  The control samples and the CG samples can be seen to
match very well in both mass and redshift (the KS-test probabilities
are $>99\%$ between CG galaxies and their controls for both mass and
redshift distributions).  The embedded and isolated galaxies are also
statistically similar to each other in both stellar mass and redshift,
with KS-test values of 32.57\% and 49.16\% respectively.  

\section{Offsets in the SFR and Metallicity of Compact Group Galaxies}
\label{sec:offsets}

In this section, we present our methodology for finding and
quantifying differences between the SFRs and metallicities of compact
group galaxies and their controls.   We also
present a set of bootstrap simulations to quantify the significance of
a given measured offset.

\subsection{Offset methodology}\label{sec:offset}

\begin{figure}
\centerline{\rotatebox{0}{\resizebox{9cm}{!}
{\includegraphics{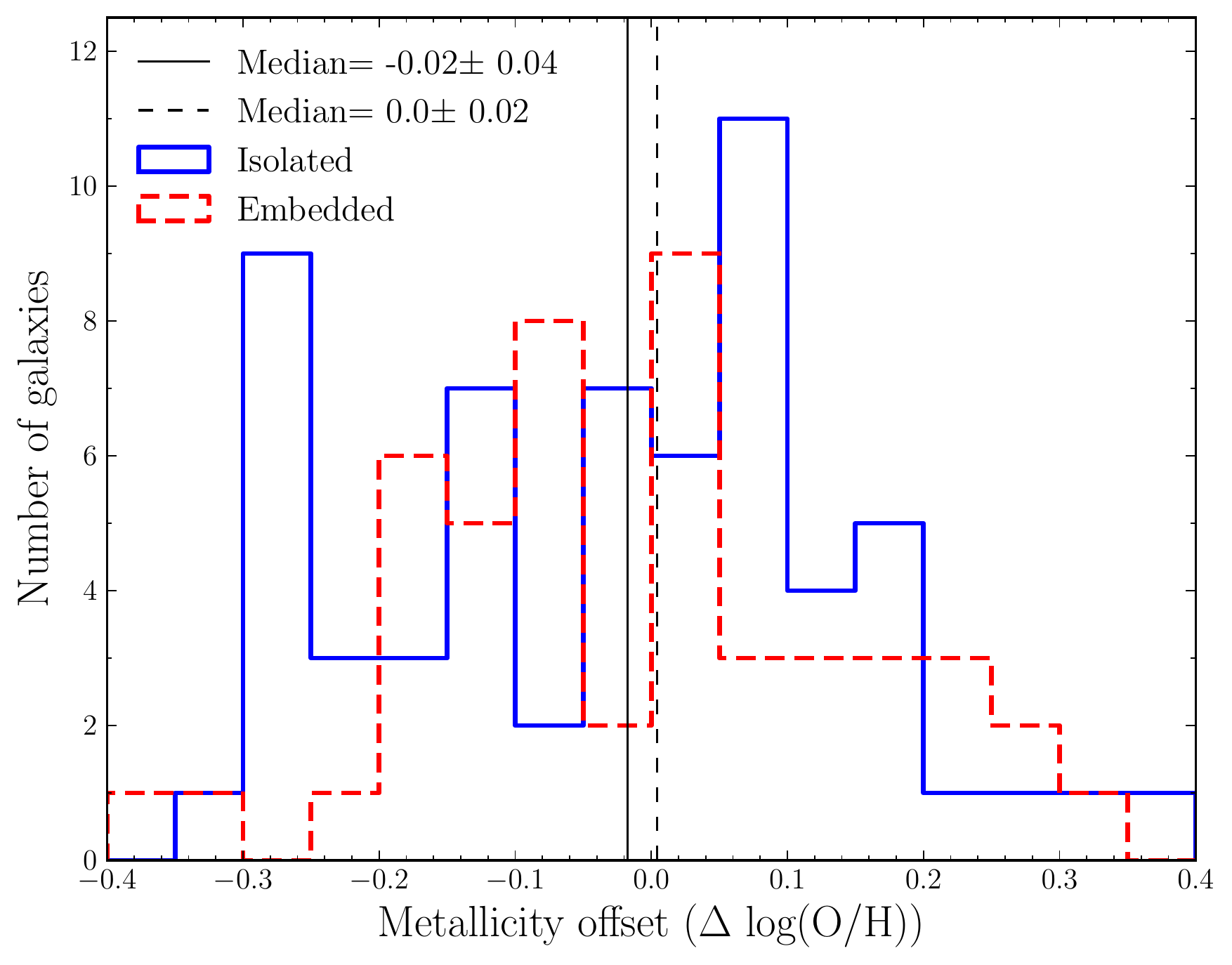}}}}
\caption{\label{fig:mzr_offset} Offsets from the control MZR as defined by the median value per set of control galaxies for the embedded (red dashed line) and isolated (blue solid)  compact group sample (50 and 62 galaxies, respectively). Black solid and dashed lines show the median offsets for the isolated and embedded compact group galaxies.  Embedded galaxies show no median metallicity enrichment, contrasted with the median metallicity offset of isolated groups, which show a dilution of $-$0.02$\pm$0.04 dex compared to the overall relation.}
\end{figure}

\begin{figure}
\centerline{\rotatebox{0}{\resizebox{9cm}{!}
{\includegraphics{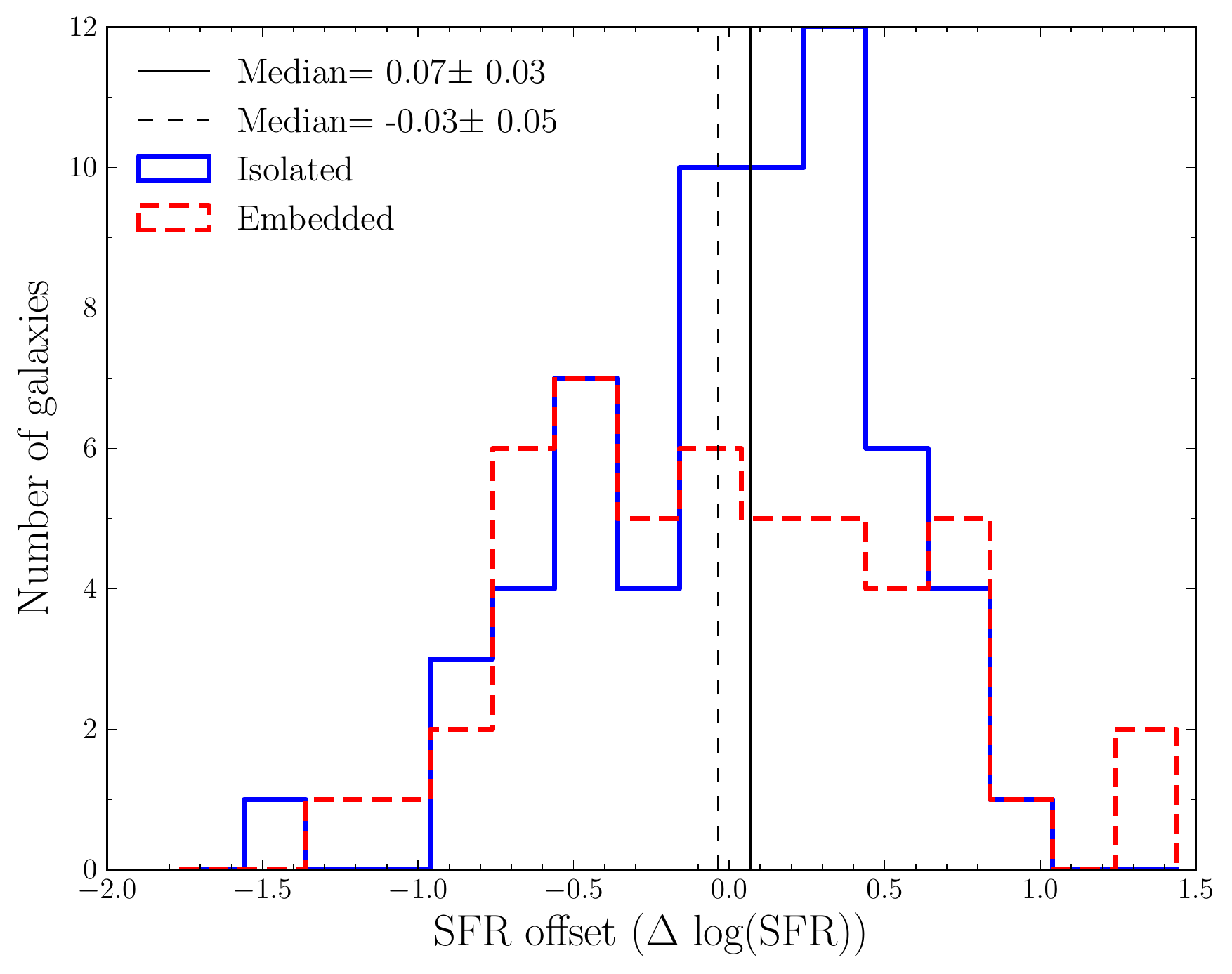}}}}
\caption{\label{fig:sfr_offset}Same as Figure \ref{fig:mzr_offset}, but now showing the offset of CG galaxies from the SFR $-$ mass relation.  Isolated CGs show a median SFR enhancement of +0.07$\pm$0.03 dex, while embedded CG galaxies have SFRs depressed by  $-$0.03$\pm$0.05 dex below the control sample.} 
\end{figure}

To quantify shifts away from the expected metallicities and SFRs at a given mass,
we use a modification of the offset method from \citet{Patton2011}.
For each galaxy in the CG sample, we take the median (metallicity or
SFR) value of all 50 control galaxies to which it was matched, and
subtract that value from the observed value of the CG galaxy. An
offset is therefore defined as:

\begin{equation}
\Delta \mathrm{log(O/H)} = \left(\mathrm{log(O/H)} + 12\right)_{observed} - \mu_{1/2}\left( \mathrm{log(O/H)} + 12\right)_{controls},
\end{equation}
\begin{equation}
\Delta \mathrm{log(SFR)} = \mathrm{log(SFR)}_{observed} - \mu_{1/2}\mathrm{log(SFR)}_{controls},
\end{equation}
~\\ where $\mu_{1/2}$ signifies the median.  The calculated offsets
are presented in Figures \ref{fig:mzr_offset} and
\ref{fig:sfr_offset}.  The median metallicity offset for the isolated
CG galaxies (solid vertical line in Figure \ref{fig:mzr_offset}) is
$-$0.02$\pm$0.04 dex (metal poor), whereas the embedded population has
a median metallicity offset of 0.00$\pm$0.02 dex (no offset; dashed vertical line).  The
uncertainty on the median is calculated using a jackknife technique.  For
each sample (embedded and isolated) the median offset is re-calculated
by systematically removing one galaxy. For a sample of $N$ galaxies,
the jackknife error on the median is given by 

\begin{equation}
\label{eq:jackknife}
\sigma = \sqrt{\frac{N-1}{N}\times\sum^N{\left(\mu_{1/2}(N) - \mu_{1/2}(N-1)\right)^2}}
\end{equation}
where $\mu_{1/2}(N)$ is the median value for the full sample of $N$ galaxies, $\mu_{1/2}(N-1)$ is the median value for $N-1$ galaxies, and each of $N$ galaxies is removed from the sample once.  

The SFR offset distribution is shown in Figure \ref{fig:sfr_offset};
the embedded CG galaxies have SFRs that are lower than the control, with
a median offset of $-$0.03$\pm$0.05 dex. Isolated CG galaxies, by
contrast, have a median offset of +0.07$\pm0.03$.  Again, uncertainties are determined from the jackknife statistic.

In order to determine whether the measured median offsets are strongly influenced by the errors on the SFRs, we use a Monte-Carlo resampling simulation to test the sensitivity of our results.  We define a distribution of possible SFR values as a gaussian with mean of the original SFR, and with width defined as the $1\sigma$ errors on the SFRs, taken from \citet{Brinchmann2004}.  Median errors on any individual galaxy are $\sim0.1$dex. We draw new SFR values for both the CG and the control samples, and recalculate the offsets for the CG samples.  We find that the median offsets are not significantly changed.  After 10,000 iterations, the median \delsfr~for isolated CG galaxies is 0.07 with a 1$\sigma$ scatter in the distribution of 0.05.  Embedded CG galaxies show a median \delsfr~of $-0.06 \pm 0.06$  (see Figure \ref{fig:mc_sfr}).  These simulated median offsets and their uncertainties are consistent with the medians in Figure \ref{fig:sfr_offset}, and we conclude that the errors on the SFR measurements are not artificially enhancing the separation between the two samples.

The uncertainties on the median offsets indicate that the metallicities of
the galaxies within both embedded and isolated CGs are consistent with a mass and
redshift-matched control sample of non-CG galaxies.  
While the median SFR of the embedded CG galaxies is consistent with the control, the median SFR of the isolated CG galaxies is significantly offset from its control at $\sim2\sigma$.   Further, the isolated and embedded medians are significantly different from each other at the $1\sigma$ level.

\subsection{Significance simulations for the SFRs}
\label{sec:sig}

The jackknife technique estimates the statistical error on our samples of $\sim60$ galaxies.  However, we also want to quantify how often a given
median offset will occur by chance (due to the intrinsic scatter
in the SFR-mass relation). A random sample of 60
galaxies is therefore bootstrapped from the control sample 10,000 times.  
Recall that each control is one of 50 that has been matched to a given CG galaxy.  When one of them is selected at random, the 49 others in the set which are not selected as the `sample' are left as the `control'.  
Offsets are then calculated between the randomly selected galaxy and each of the remaining controls.  The median offset for the sample of 60 galaxies is determined, along with the jackknife error on that median.
As the galaxies selected
are part of the control pool, and no offset is artificially
introduced, the majority of these offsets should be centred around
zero.  For each median, the significance $\sigma$ is calculated by dividing the measured median offset by the jackknife error on the median.
 The fraction of the total number of runs with significant offsets (defined as $\sigma > 1,2, 3,$ or 4) can then be calculated.  

\begin{figure}
\centerline{\rotatebox{0}{\resizebox{9cm}{!}
{\includegraphics{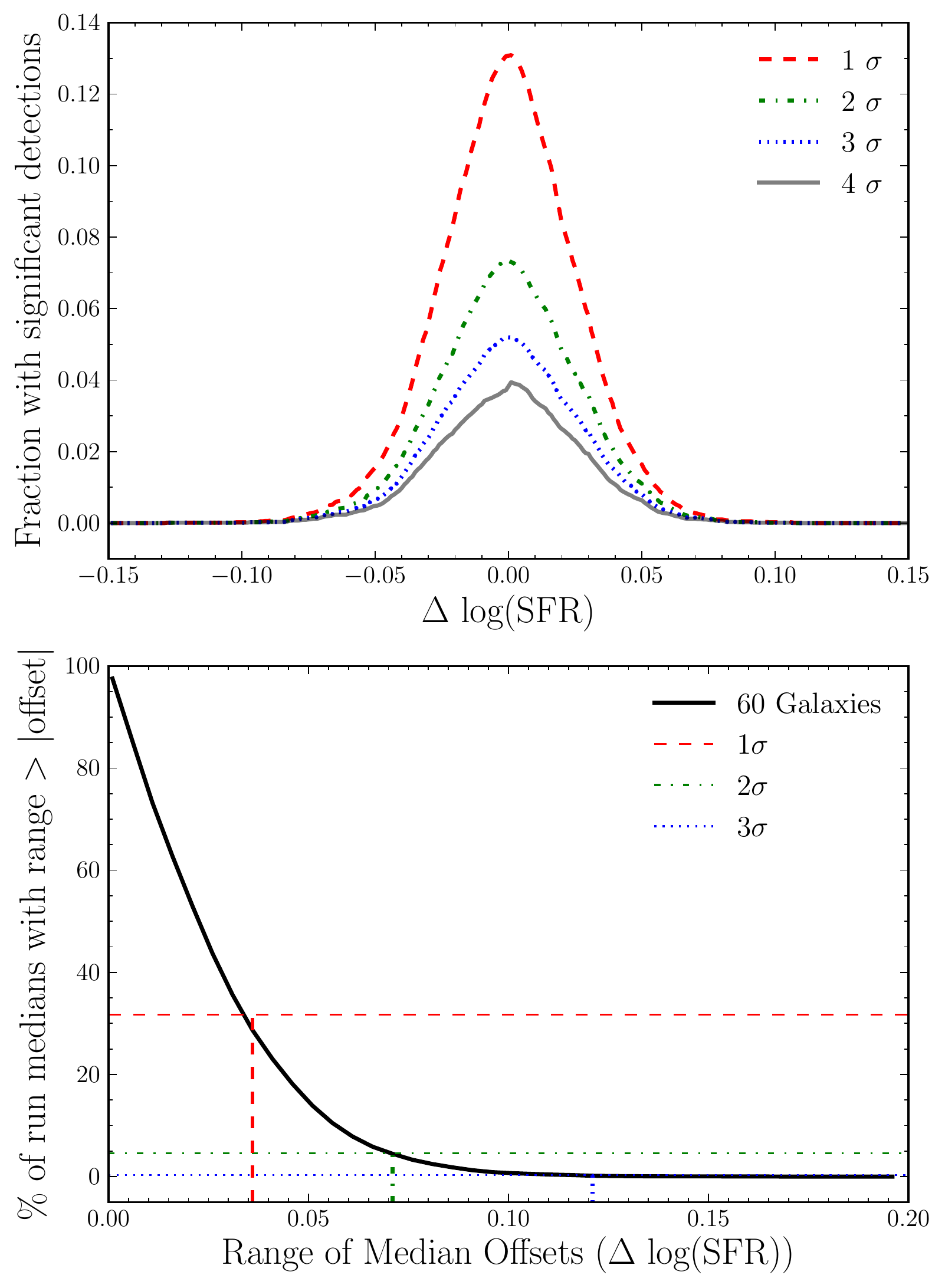}}}}  
\caption{\label{fig:boot_sfr} Results of the bootstrapping simulation for SFRs.  A randomly selected sample of 60 galaxies is pulled from the control pool, and their offsets from their matched controls recorded, repeated 10,000 times.   The top panel is the fraction of the total number of runs where a significant offset (i.e., $\sigma >1,2, 3,$ or 4) greater than a given offset $i$ is found.  Offsets of $-0.04$ and $+0.08$ dex at 2 and 4$\sigma$ significance have $ 2.14$\% and 0.02 \% likelihood of arising by chance, respectively.  The bottom panel tests the percentage of times that a range in medians between samplings occurs; a difference in median of 0.10 dex could occur randomly $\sim$ 0.67\% of the time ($> 2\sigma$).}
\end{figure}

We can now compute for any given offset the fraction of the total number of bootstraps which returned significant medians at or beyond the offset being investigated.  The results are shown in the top panel of Figure \ref{fig:boot_sfr}.
The different lines indicate different values of $\sigma$. To compare to our results, the SFR enhancement seen in isolated CGs would be along the $2\sigma$ line.
This simulation indicates that there is a 0.19\% chance of detecting a SFR enhancement of +0.07 $\pm 0.03$ dex at random.  
We therefore conclude that the SFR enhancement we have detected, at its given observed significance, has a very low probability of occurring by chance due to scatter.

Whilst the above procedure assesses the probability that a given
offset is found between the CG and control samples, we can also
quantify the significance of the difference \textit{between} the
embedded and isolated samples.  This is achieved by repeating the
bootstrapping procedure described above, but this time selecting 2
separate samples of 60, and the absolute value of the difference in
their median offsets is found.  We then find the percentage of iterations where the simulated
differences in median is greater than a given offset.  The results of
the two sample bootstraps are shown in the lower panel of Figure
\ref{fig:boot_sfr}. 
The probability of a range of 0.10 dex between
two samples drawn at random from the control SFR-mass relation is $0.67\%$, which corresponds to $>2\sigma$ significance.  
 
\subsection{Upper limit simulations for metallicities}

\begin{figure}
\centerline{\rotatebox{0}{\resizebox{9cm}{!}
{\includegraphics{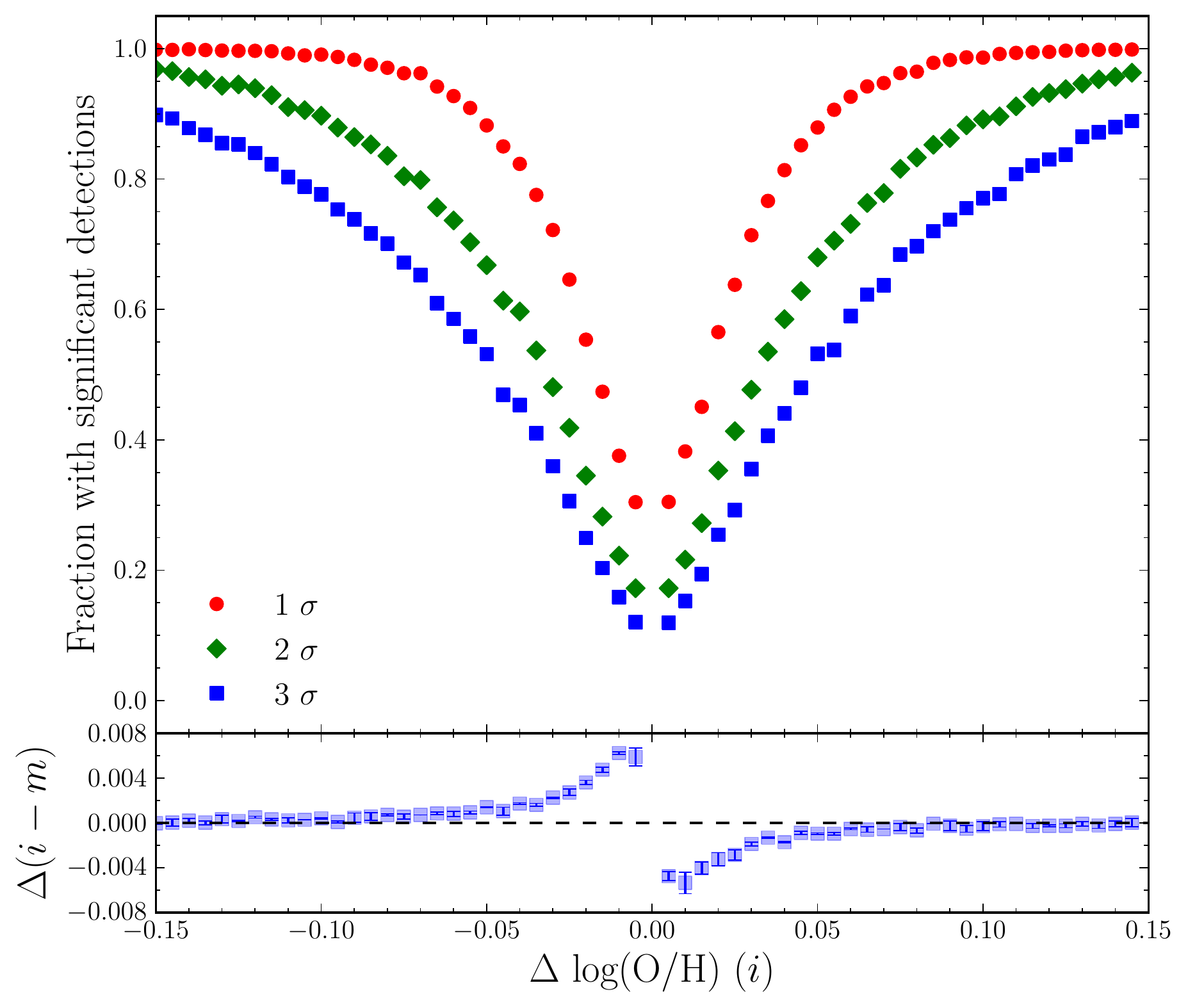}}}}
\caption{\label{fig:sim_mzr}Results of the sensitivity simulation for the MZR offsets.  Each point is the result of an independent simulation, bootstrapped 5,000 times.  Each simulation selects a random sample of 60 galaxies from the control sample, and inserts an artificial offset $i$ into the metallicities.  The offsets are then calculated as normal, resulting in a measured offset $m$.  The fraction of those runs which result in significant detections at the 1,2, and 3 $\sigma$ level are plotted in black, red, and blue, respectively.  The median difference between the measured offset $m$ and the input offset $i$ for the fraction of significant runs (at $3\sigma$) is displayed as the blue squares in the lower panel as a function of $i$.  Jackknife errors on the medians are smaller than the points.}
\end{figure}

Although we find a statistically significant difference between the
SFRs of embedded and isolated CGs, the metallicity offsets are
inconclusive ($-0.02\pm0.04$ and $0.00\pm0.02$ for isolated and
embedded CGs respectively).  Metallicity offsets in both embedded and
isolated structures are statistically consistent with their control
samples.  To test the sensitivity of our null result given the sample
size, we run a set of simulations similar to the bootstrap used in \S
\ref{sec:sig}, but now introducing an artificial offset.  In this
simulation, 60 galaxies are randomly selected from the control sample
and an artificial, single value offset ($i$) is added to all the
metallicities of the selected galaxies.  The resultant metallicities
are then run through the offset calculation algorithm.  Offsets are
calculated between each of the 60 randomly selected galaxies and the
remaining 49 in its matched set.  The median measured offset ($m$) of
the 60 galaxies is recorded.  This step allows us to quantify the
difference between the input offset $i$ and the recovered offset $m$
through our algorithm.  We expect $i-m=0$ if the sample can accurately
recover the input metallicity offset. This bootstrap process is
repeated 5,000 times.  We calculate the jackknife error on each of the
median values returned by the bootstrap.  For each realisation, we
calculate a significance $\sigma$ by dividing the measured offset $m$
by the jackknife error.  We can then find the fraction of the total
number of bootstrapped samples which return significant offsets for a
given $i$.  This fraction is then recorded, and plotted in the top
panel of Figure \ref{fig:sim_mzr}.  Red, green, and blue points
reflect the fractions for $\sigma > 1,2,$ and 3, respectively.  Each
inserted offset requires a separate, independent run of 5,000
iterations of the bootstrap, and is plotted as a separate point in the
figure.

In the bottom panel of Figure \ref{fig:sim_mzr}, the median difference between the input and measured offsets ($i-m$) is shown for those runs which returned significant medians at the 3$\sigma$ level, as a function of the inserted offset $i$.  This panel indicates that we are able to accurately recover an offset, if it is systematically different from the control.  For values of $i$ smaller than $\pm0.05$, $(i-m)$ shows an asymmetric shift away from zero.   At very small $i$, it becomes increasingly unlikely for an offset $m$ to be measured as significant.  Therefore, those simulations which are measured as significant at small $i$ are likely to be those which have scattered to median values further from zero.  Negative values of $i$ therefore preferentially return more negative values of $m$, and positive values of $i$ return more positive values of $m$.

With these simulations as a base, we may compare the fractions of
significant measured offsets returned by any given iteration of the
bootstrap with the magnitude of offset we might expect to see.  We use
what has been found in previous studies of metallicity changes in high
density environments as a guideline for reasonable magnitudes of
offsets.  The metallicity suppressions found in the pairs sample of
\citet{Ellison2008} and of \citet{Michel-Dansac2008} are of order 0.05
dex, and the enhancements in overdensities in \citet{Cooper2008} are
$\sim$0.03 dex.  Similar magnitudes are found by \citet{Ellison2009},
who find offsets of 0.04 dex in galaxies in high density regions,
relative to their control.  \citet{Mouhcine2007} find slightly higher
differences in metallicity of 0.06--0.08 dex between high and low
density environments.  For offsets in the range 0.03 -- 0.05 dex, we
find we only have a 34.8--51.1\% chance of detecting a 3$\sigma$
result.  We do not reach 99\% confidence levels of detecting a
1$\sigma$ offset with our sample size until we reach offsets of 0.13
dex, which is much larger than anything previously discussed in the
literature.  The lack of significant metallicity offset in our CG
sample may therefore be a limitation of our sample size.  The fact
that our metallicity results are both consistent with the control,
combined with the results of this simulation, indicate our sample is
unlikely to result in a significant offset of the magnitude suggested
by the literature. 

In summary, the offsets in SFR in our CG samples appear to be
statistically distinct from each other, and galaxies in isolated CGs have significantly enhanced SFRs, whereas embedded CGs show no enhancement.
However, the measured metallicity offsets are
not sufficiently sensitive to probe metallicity differences between
embedded and isolated compact group galaxies and their controls.

\section{Discussion}
\label{sec:discussion}
We have used the varying large scale environments of a consistently selected sample of compact, high density groups to probe the influence of the presence of a rich group or cluster.   
The compact group selection criteria allow us to select a consistently dense local environment on scales of $\sim100$ kpc (see Figure 6 of M11) which, ignoring all external influences, ought to be subject to the same internal evolutionary processes.  Galaxies in CG systems
typically have small projected separations from each other; for our sample the median projected separation to the nearest neighbour is 46 kpc for both isolated and embedded samples. In galaxy pairs separations $\lesssim$ 30 kpc result in significantly  enhanced SFRs \citep[e.g.,][]{Barton2000,
Lambas2003, Alonso2004, Alonso2006, Ellison2008, Ellison2010, Scudder2012b}.  Our sample has a significant population of galaxies with separations of this order, making galaxy--galaxy interactions an appealing explanation for the SFR enhancement seen in the isolated CG sample.  
The level of offset seen in our sample of isolated CG galaxies (+0.07 dex) may appear modest, but is comparable to studies of merger-induced star formation, which generally find median SFR enhancements of around +0.1 dex \citep[e.g.,][]{Ellison2008, Scudder2012b}.  
If the isolation criterion of the CG selection is effective at limiting
outside influences, and interactions are the only dominant mechanism
for transformations, we might expect both samples of CG galaxies to show consistent
SFR enhancements over the control.

We find that galaxies in CGs in close
proximity to a cluster or rich group ($<$1 Mpc) do not show
the same preferential enhancement as their counterparts in isolated CGs.
The embedded CG galaxies instead show consistency with the average galaxy in the control.
Galaxies in isolated CGs ($>$1 Mpc from large scale structures), by contrast, show SFR
enhancements over the control, consistent with the general trend presented by the SFR-density relation.  
This SFR disparity indicates that the
local overdensities identified by the CG selection criteria cannot be
considered in isolation.  The presence of a nearby rich group exerts
$some$ influence on the star forming galaxies found within CGs, either by
introducing an additional quenching mechanism, or because the galaxies
within rich groups are physically distinct from their isolated
counterparts. The SFRs and metallicities of the galaxies within those groups therefore serve as diagnostics of the impact of the larger scale environment.   
\textit{However, given the small magnitude of the difference between the two CG samples, the large scale environment does not appear to drive major changes in a galaxy's evolution at fixed local density.}

\subsection{Density scale dependences}
 Having determined the sensitivity and significance of the metallicity and SFR offsets, we wish to compare these results to previous work.   However, previous metrics of environment usually fall into one of two categories; either an $n^{th}$ nearest neighbour approach, or calculating galaxy density within annuli of increasing radius around a given point.
In this work, we have used a new method of determining the large scale environment of a galaxy.  
Our division of galaxies into embedded and isolated systems is a binary assignment, based upon the presence (or lack) of rich group structure within a cylinder in redshift and projected separation space, centred upon the CG.  This is an entirely different method of characterising the nature of a galaxy's environment than has been typically used in the past.    

Comparisons to the results found using existing measures of environment will require attention to the methodology of those metrics to be sure that they can probe the same kinds of overdensities.  
Since the $n^{th}$ neighbour and annulus methods are calculated using entirely different algorithms, the meaning of the densities returned (as it applies to our CG sample) will vary, and comparisons to the results presented here must bear this in mind.  In addition, most previous work has focused upon the \textit{relative} importance of local and large scale dependences.  
This relative comparison is one that we cannot address in this work, as our classifications of embedded and isolated are not relative selections, and the local density - that of the CG itself - is relatively constant across our sample. 
Therefore, in comparing to previous work, we can only determine whether or not the previously obtained results are consistent with some level of SFR dependence on large scale environment as measured by our own method, to the extent that the density metrics may be compared.

The $n^{th}$ nearest neighbour approach identifies high density regions of space by returning a shorter distance to the $n^{th}$ neighbour in dense regions, and a longer distance in lower density regions \citep[e.g.,][]{Hashimoto1998, Carter2001}.  Varying the value of $n$ effectively changes the distance scales over which one wishes to probe the densities.  With small values of $n$, the densities returned will be more reflective of local scale densities.  Larger values of $n$ will necessarily wander further afield to reach more distant neighbours.
These distant neighbours will return a much larger volume of space over which to average the galaxy density, relative to their locally dense counterparts.  
The larger distances for isolated systems means that the densities derived from the identified $n^{th}$ neighbour will be more reflective of a large scale density than those in locally dense environments (which return smaller distances), even if those local overdensities are relatively isolated from other structures \citep{Weinmann2006a}.   Since the distance to $n^{th}$ nearest neighbour can vary smoothly from system to system, the distance scales being probed will also vary strongly as a function of local galaxy density.  Furthermore, since the distance to  $n^{th}$ nearest neighbour is converted into a density, information about the distribution of those galaxies on scales less than that distance is lost.
For the purposes of comparison to our metric, the $n^{th}$ nearest neighbour algorithm simply needs to be able to distinguish between isolated and embedded CG systems.  Both isolated and embedded CGs have similar densities in the cores, but exhibit different galaxy surface densities on scales of 0.1 -- 1 Mpc (M11).   As long as the  distances returned by the $n^{th}$ nearest neighbour probe these scales, they should be able to distinguish between the isolated and embedded systems.

This relatively straightforward criterion for comparison with the $n^{th}$ nearest neighbour provides a sharp contrast with the other main metric of galaxy density.  The annulus approach is more difficult to compare to our method.  In this approach, the galaxy density within annuli of increasing radius is measured \citep[e.g.,][]{Lewis2002, Kauffmann2004, Blanton2006}. Many works place their dividing line between `local' and `large' scale at 1 Mpc.  The densities within the annulus of radius 1--3 or 1--6 Mpc is therefore used as their measure of large scale environment.
1 Mpc is also a natural dividing line in our sample, but only in the sense that this is the distance to the nearest group that distinguishes isolated and embedded CGs.
The annulus approach is sensitive to the presence of broad overdensities beyond the distance of the defined `large scale', generally 1--6 Mpc away from the selected galaxy.  Our method provides the ability to define a system as embedded even when the overdensity is entirely on scales $<$ 1 Mpc.  In this case, the CG would still distinctly be embedded in a more massive halo than its own, but would not be distinguished as `large scale' structure by an annulus of radius $>$1 Mpc.  If some fraction of the rich group extends beyond 1 Mpc distant from the CG galaxy, then it would be detected by the $>$1 Mpc annulus.  However, M11 shows that rich groups and CGs in any environment converge to similar galaxy densities at separations $\sim$ 1 Mpc.  With the median distance from CG centre to a rich group for embedded groups being of order $\sim0.3$ Mpc, and the majority of rich groups's strong galaxy surface density visible on scales $<$ 1 Mpc, we might expect that only a small fraction of the galaxies within a rich group to extend beyond the 1 Mpc threshold.  Furthermore, the annulus method loses all information on scales smaller than the area of the annulus.  If an annulus contains an overdensity, but also contains a very low density region, these two extremes will average to a value that is not representative of either region.

With these differences in mind, we can now attempt to compare our results of a SFR dependence on large scale environment with previous studies.  We emphasize that we are investigating only the dependence on those galaxies which are star forming relative to a star forming control, not the overall trend of star formation rate with density.  Those works which conclude that the large scale environment affects the evolution of star forming galaxies by way of the cluster-centric radius \citep{Park2009a} or dark matter halo mass \citep{Goto2003, Weinmann2006a} are compatible with our results.  However, a large body of work in the literature (using a mixture of methodologies) has concluded that the local scale effects dominate galaxy evolution \citep[e.g.,][]{Hashimoto1998, Carter2001, Lewis2002, Kauffmann2004, Blanton2006, Blanton2007}.  These analyses consistently find a strong SFR dependence upon local galaxy density, with either very weak or absent dependence upon galaxy densities beyond the 1 Mpc scale.  
For example, \citet{Blanton2006} constrains the effects of large scale (1-6 Mpc) effects to be small, but does not exclude them as a minor contributor.  
Our results are not in disagreement with those studies which found that the local scales dominate galaxy evolution, consigning the large scale effects to a minor role. 
\textit{The small difference in \delsfr~ between star forming galaxies in isolated and embedded CGs surely confirms the minor nature of the large scale environment.  Our work indicates simply that there is a non-zero effect upon the galaxies within a larger structure beyond local density effects; it does not require that the effect must be the dominant one.}
Other works in the literature are less conservative, and interpret the lack of a strong dependence upon the galaxy densities at $>$ 1 Mpc radii as evidence that the large scale environment does not have an effect upon a galaxy.  \citet{Blanton2007} find that for the star forming fraction, large scale structure (defined as structure in an annulus $> 1$ Mpc) has no discernable effect upon the galaxy, with a bias in their measurement of no more than 5\%, and density error of order 15\%.  \citet{Kauffmann2004} found that local densities dominate the SFR evolution; at low masses, the SFR of a galaxy was up to 10 times lower in high local densities than it was in low local densities. When testing the spatial scales between 1 -- 3 Mpc, they found a very noisy relation in the 4000 \AA ~break strength, their metric for SFR, with density: consistent with no dependence.  In contrast, we have found that SFRs do depend, albeit at a low level, on their association with a large scale structure.

Reconciling our results with those in the literature likely comes down to the nature of the `environment' identified by the different density metrics that have been previously used. 
 Our division in environment identifies galaxies which are within 1 Mpc of a larger group than the CG itself.  By the time scales of $>$ 1 Mpc from the centre of a group are reached, the galaxy densities for isolated, embedded, and rich groups have all converged (M11).   This indicates that most of the galaxy overdensity due to the rich group has also dwindled by 1 Mpc from the centre.  If the groups are then sufficiently far away from the CG, they might cross over into the $>$ 1 Mpc annulus, and be detected as an overdensity by that metric.  However, the peak of the distribution of the distance to the nearest rich group is $\sim$0.3 Mpc for embedded structures (M11).  At 0.3 Mpc, the majority of the rich group will be found within 1 Mpc of the CG, and would not significantly increase the measured density for annuli $>$1 Mpc.  Therefore, even though the embedded CGs are situated in a significantly different environment than the isolated CGs, the overall galaxy density no longer reflects this difference at scales $>$ 1 Mpc. Dividing environmental factors into large and local scale effects simply by using a cut in radius may blend extremely local drivers of evolution and some larger scale environmental factors.
Our metric therefore cleanly separates those groups embedded in a more massive halo from those which are independent halos, no matter what the projected scale of the more massive group may be.

\subsection{Physical Drivers}
The interpretation of the difference between embedded and isolated galaxy SFRs depends on whether the galaxies are intrinsically and physically different when they are found within an overdensity, relative to in isolation, or whether the overdensity is imposing an additional processing mechanism on an initially similar set of galaxies.  If the former, then galaxies found within large scale structure are reacting to the same set of evolutionary processes as galaxies in isolation, but their intrinsic properties make SFR enhancement (for example) more difficult to achieve.  

One mechanism by which a physical difference in the galaxy populations between embedded and isolated CGs might arise is assembly bias \citep[e.g.,][]{Croton2007,Cooper2010}.  Assembly bias states that galaxies in high mass potentials tend to be older, as they collapsed and began to cluster at earlier times.  In this scenario, one might expect clusters to contain a higher fraction of old, so-called `red and dead' galaxies, galaxies which have undergone many mergers, or galaxies which have been affected by the cluster potential through quenching mechanisms for a longer period of time.  Galaxies in rich groups tend to be preferentially gas poor \citep[e.g.,][]{Chung2009} as a result.  This framework offers one explanation for the SFR suppression in overdense regions.  
If galaxies within rich groups are simply older than their counterparts in the isolated groups, then they would be further along in their evolution, and may be further along the process of gas exhaustion and consumption.  If the CG galaxies, due to their age, in addition to any environmental effects, are preferentially gas poor, then it will be significantly more difficult to trigger a SFR enhancement after an interaction \citep[e.g.,][]{Alonso2006}. 
Their isolated counterparts, having collapsed at later times, are younger, more gas-rich, and at a much earlier stage in their evolution; these galaxies will show SFR enhancements much more easily, as their gas reservoir is less depleted.  
This explanation does not require a difference in the mechanisms that modulate star formation rates in the two environments; the difference in intrinsic galaxy properties is enough to alter the galaxy's response to the same large scale environmental mechanisms.  
However, all the galaxies in our sample are emission line galaxies.  While their star formation rates may be lower than the average galaxy, they still have significant amounts of gas.  They cannot have been so affected by quenching and stripping mechanisms as to remove all their gas, if they can maintain strong emission lines.  A large portion of the galaxies in our sample, regardless of their classification as embedded or isolated, retain relatively late-type morphologies (see Figures \ref{fig:emb_mosaic} \& \ref{fig:iso_mosaic}).  The maintenance of the emission lines and their late-type morphologies limits the magnitude of the environmental and age-related evolution that the galaxies could have undergone.

Turning to consider processes that could alter the SFRs of a given galaxy in a different environment, we can eliminate quenching mechanisms that function on very small scales such as those associated with interactions; these should be consistently affecting both isolated and embedded systems.  Whatever mechanism is preventing SFR enhancement in embedded CG galaxies must be associated with the larger scale cluster potential, and not simply with the high densities of the CG system.  One explanation could be the difference in evolutionary processes in central and satellite galaxies.  Galaxy populations divided in this way show that galaxies which are not the dominant galaxy in the dark matter halo undergo stronger environmental processes than the central galaxy does \citep{Haines2006, Kimm2009, Peng2011, Tinker2011}.  Embedded CGs would, by definition, be a satellite halo within the larger cluster halo.  The enhanced processing seen in most satellites would therefore be systematically applied to all galaxies in the embedded systems.  \citet{Blanton2007} find that high density substructures (on scales of $<300$ kpc) within a cluster are more correlated than expected with hosting red galaxies, i.e., these galaxies are further along in the quenching process.  By contrast, the probability of the dominant galaxy in the potential being in our sample is significantly higher for isolated systems, where the CG environment is more likely to be an isolated halo.  

If the difference between galaxies in isolated and embedded CG systems is not due to a systematic physical difference in galaxy population, then the galaxies in both systems ought to have begun their evolution in the CG as a roughly uniform population of galaxies.
In order to explain the absence of the SFR enhancement seen in the isolated CG galaxies, an additional evolutionary force must be at work in embedded CG galaxies due to the cluster potential that is not in effect in isolated halos.  This brings into suspicion the main drivers of galaxy evolution in clusters: ram-pressure stripping, galaxy-galaxy harassment, and strangulation.  The isolation criterion for CG selection means that external galaxy harassment is unlikely to be a significant influence.  Galaxy harassment within the CG would contribute similar levels of evolutionary processing to both galaxy populations of CGs, and is therefore not a useful explanation for the difference in SFR response.  Further, CGs have relatively low velocity differences, so harassment within the group itself is unlikely to dominate over lower velocity tidal interactions.
This leaves ram-pressure stripping (removal of cold gas from the disk) and strangulation (removal of the external gas supply).  
\citet{Weinmann2006a} argue that ram-pressure stripping should affect low mass galaxies more quickly, as they have a smaller potential and are more easily stripped of their gas, within a halo of given mass.  This ought to result in a higher fraction of low mass early type galaxies relative to the fraction of high mass early types.
Their sample, taken from the SDSS DR2, contradicts this prediction, finding no evidence for preferential changes to the lower mass galaxies. 
\citet{Wetzel2011} find that galaxies introduced into a larger halo can retain their SFR at pre-accretion levels for 1-2 Gyr before rapid quenching takes place; if ram-pressure stripping were the dominant mechanism, the galaxies would have their cold gas removed on much quicker timescales. Ram-pressure stripping is also excluded by a number of other authors \citep[e.g.,][]{Tanaka2004, Blanton2007} due to the large radial extent of SFR suppressions, and SFR suppressions in low mass halos.  Neither of these environments are expected to have the dense intra-cluster medium needed to support effective ram-pressure stripping \citep[e.g.,][]{Rasmussen2008}.
This leaves strangulation as the prime suspect for SFR suppression in cluster environments, as it could still remove the hot halo of gas around a galaxy in lower density or lower mass environments \citep{Balogh2000}.  Furthermore, it does not affect the morphologies or cold gas reservoirs of the galaxies involved \citep{Weinmann2006a}.  M11 find that the morphologies within embedded and isolated CGs are very similar, which also supports the idea of an environmental process being at work, rather than different evolutions due to age or gas exhaustion.    Strangulation therefore appears to be an appealing mechanism for long-term suppression of SFR in emission line galaxies.

\section{Conclusions}
\label{sec:conclusions}
We present a sample of robustly calculated gas-phase metallicities and
star formation rates for a sample of 75,863 star-forming galaxies in
the SDSS DR7.  These metallicities use emission line fluxes which
passed rigorous quality control, and are cleaned of any AGN
contribution.  From this sample, we define a compact group sample of
112 galaxies that is uniformly selected on criteria for local richness
and cleaned of likely interlopers.  The compact group sample is
further split by large scale environment into isolated and embedded
subsamples, with 62 and 50 galaxies in each sample respectively.
These two samples of star forming CG galaxies therefore have the same (very high)
local density, but different large scale environments.  The samples of
compact group galaxies are simultaneously matched in mass and redshift
with 50 non-compact group galaxies each in order to form robust control
samples.  By comparing the control samples with the compact group
galaxy properties, we can measure the difference in the SFR and
metallicity at fixed mass. The main conclusions of our work are as
follows:

\begin{enumerate}
\item{Isolated and embedded compact group galaxies are offset from each other, and isolated CG galaxies are significantly enhanced relative to the field SFR-mass relation.  Embedded galaxies do not show a similar enhancement.  The SFRs of the galaxies in isolated CG systems show a median enhancement of +0.07$ \pm $0.03 dex, whereas the galaxies in embedded systems show a median offset of  $-$0.03$ \pm $0.05 dex.  A series of bootstrap and Monte Carlo simulations indicate that this offset is robust.  Our results indicate that, even though previous works have found a primary dependence on local density, the SFRs of star forming galaxies in locally overdense regions are mildly sensitive to large scale environment at fixed local density. }

\item{We find no evidence for a significant metallicity offset in either the isolated or embedded samples relative to their control samples.  
However, this non-detection is likely to be a result of small number statistics, as our sample is not likely to detect offsets of order 0.03--0.05 dex, which is the typical metallicity offset expected due to interactions or density driven effects.  Our sample is therefore inconclusive with regards to the metallicity offsets.  The simulations indicate that at 99\% confidence we could have detected an offset of 0.13 dex.}
\end{enumerate}

Criteria classically used to select CGs define a consistent set of high density groups of galaxies.  However, when these systems are divided according to the presence of a nearby cluster or rich group, the SFRs of the star forming galaxies within these systems become divided.  The environmental split in median SFR demonstrates that large scale structure can exert an effect upon substructure found within it, even if at a low level, and that substructure is relatively isolated from other galaxies.
The two possible explanations are: galaxies found within overdense regions are intrinsically different from galaxies in low density regions, which alters their response to the same local overdensities, or that the rich group structure imposes an additional evolutionary mechanism onto the galaxies in embedded CGs.  
This may give further support to the differences in satellite and central galaxy evolution, with strangulation, found only in cluster environments,  as an effective process in slowly shutting down star formation.

 \section*{Acknowledgments} 
We thank the referee for useful comments which improved this paper.  We also thank David Patton, Michael Balogh, Chris Pritchet, Alan McConnachie, and Bianca Poggianti for their comments on a draft of this paper.

We are grateful to the MPA/JHU group for access to their data products and catalogues (maintained by Jarle Brinchmann at  \url{http://www.mpa-garching.mpg.de/SDSS/}).

Funding for the SDSS and SDSS-II has been provided by the Alfred P. Sloan Foundation, the Participating Institutions, the National Science Foundation, the U.S. Department of Energy, the National Aeronautics and Space Administration, the Japanese Monbukagakusho, the Max Planck Society, and the Higher Education Funding Council for England. The SDSS Web Site is \url{http://www.sdss.org/}.

The SDSS is managed by the Astrophysical Research Consortium for the Participating Institutions. The Participating Institutions are the American Museum of Natural History, Astrophysical Institute Potsdam, University of Basel, University of Cambridge, Case Western Reserve University, University of Chicago, Drexel University, Fermilab, the Institute for Advanced Study, the Japan Participation Group, Johns Hopkins University, the Joint Institute for Nuclear Astrophysics, the Kavli Institute for Particle Astrophysics and Cosmology, the Korean Scientist Group, the Chinese Academy of Sciences (LAMOST), Los Alamos National Laboratory, the Max-Planck-Institute for Astronomy (MPIA), the Max-Planck-Institute for Astrophysics (MPA), New Mexico State University, Ohio State University, University of Pittsburgh, University of Portsmouth, Princeton University, the United States Naval Observatory, and the University of Washington.
\bibliographystyle{apj}
\bibliography{masterfile_bibdesk}
\begin{appendix}
\section{Figures \& Tables}
\begin{figure*}
\centerline{\rotatebox{0}{\resizebox{13cm}{!}
{\includegraphics{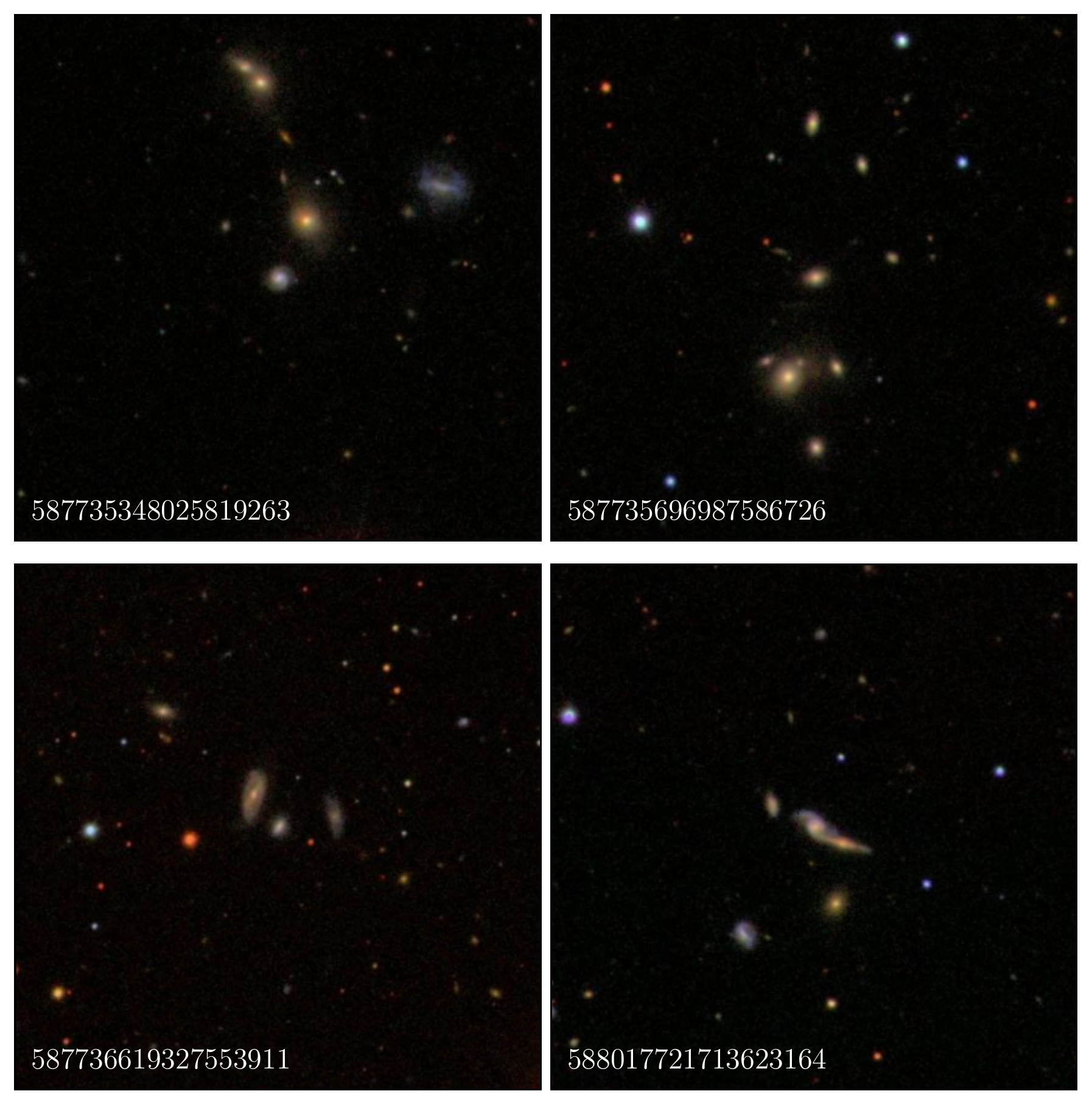}}}}
\caption{\label{fig:emb_mosaic}4 embedded CG galaxies, selected at random.  The images are centred on the galaxy in the spectroscopic sample (identified by SDSS objID in the lower left corner), and are approximately 200 arcseconds to a side.}
\end{figure*}

\begin{figure*}
\centerline{\rotatebox{0}{\resizebox{13cm}{!}
{\includegraphics{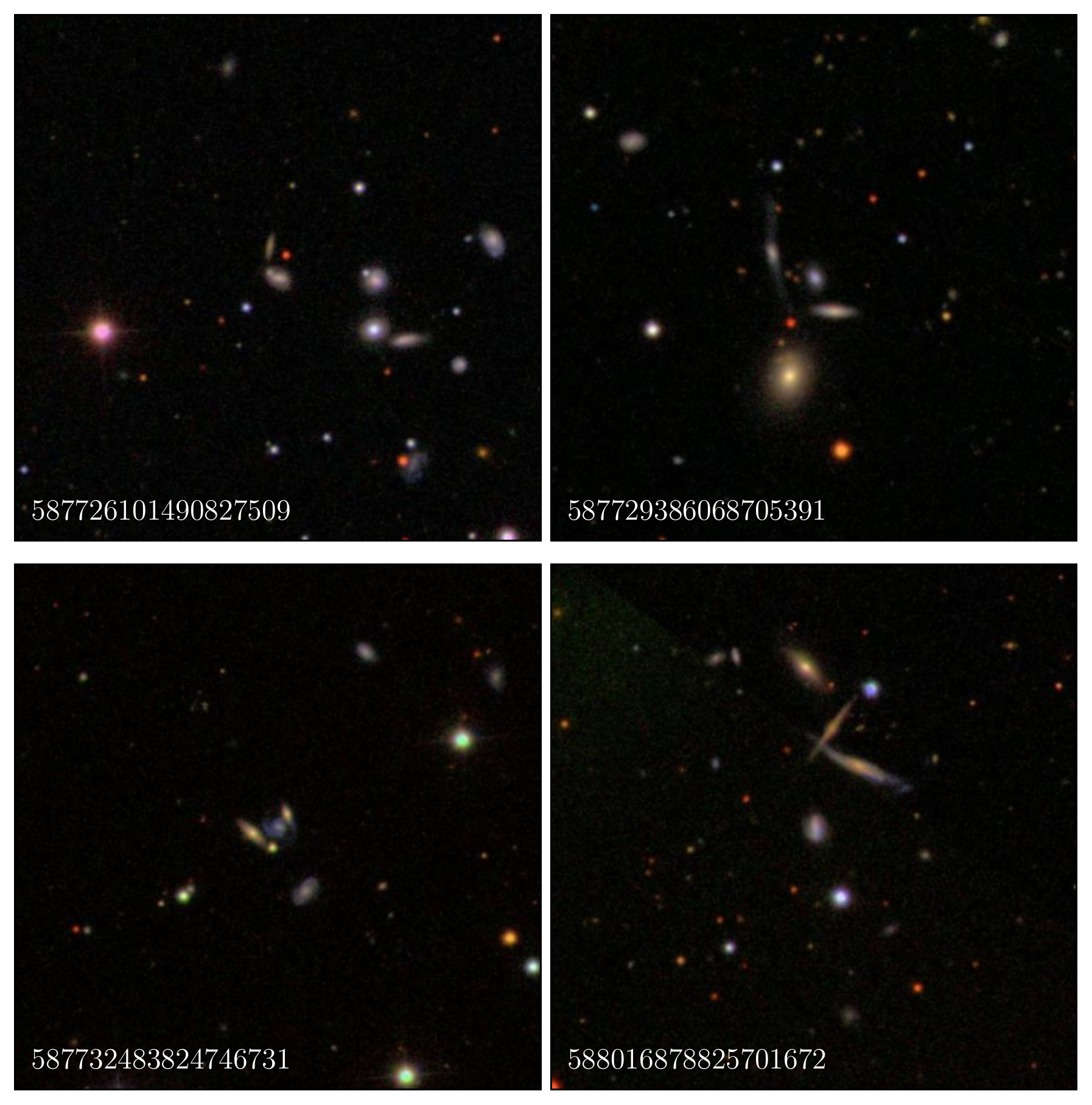}}}}
\caption{\label{fig:iso_mosaic}4 isolated CG galaxies, selected at random.  The images are centred on the galaxy contained in the spectroscopic sample (identified by SDSS objID in the lower left corner), and are approximately 200 arcseconds to a side.}
\end{figure*}

\begin{figure}
\centerline{\rotatebox{0}{\resizebox{7.15cm}{!}
{\includegraphics{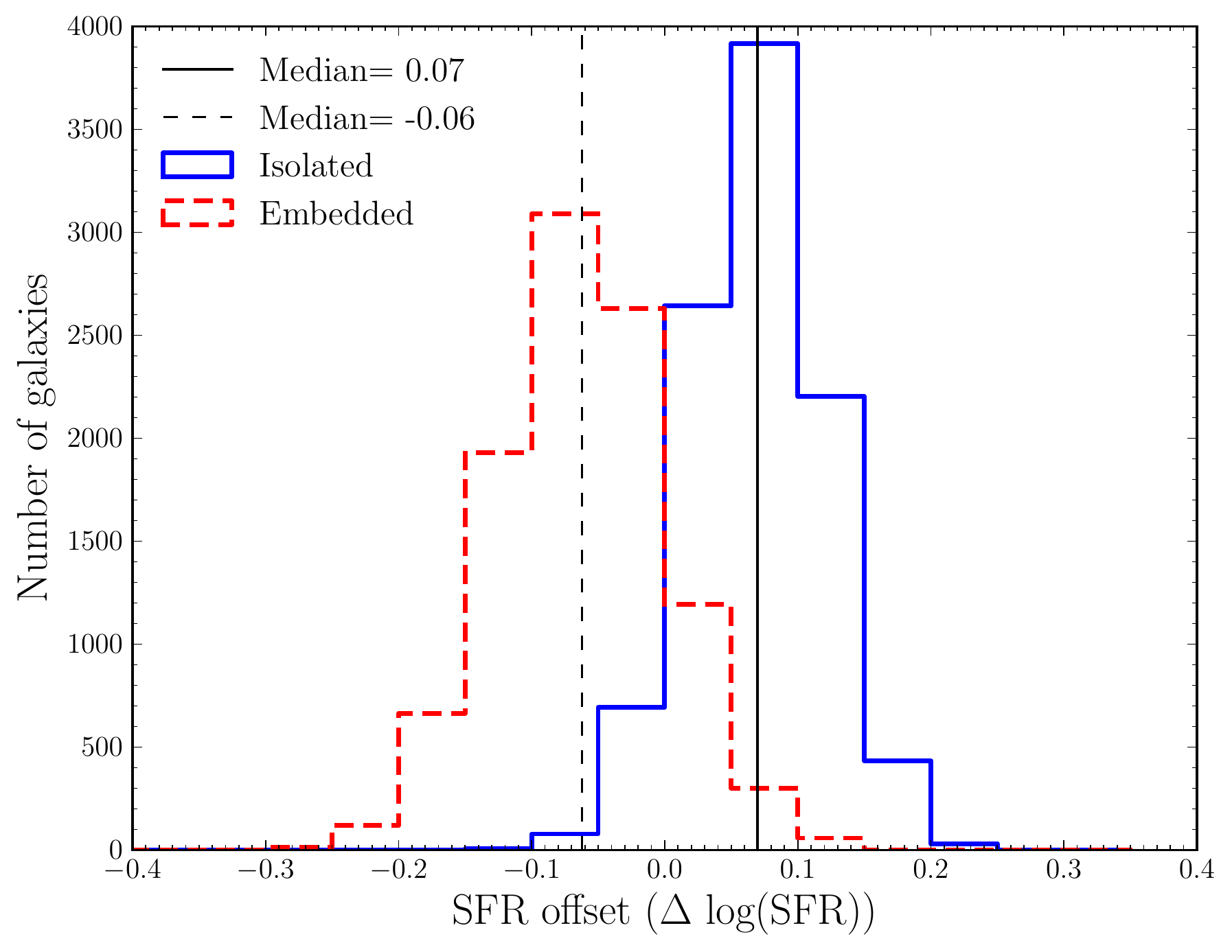}}}}
\caption{\label{fig:mc_sfr}Results of a Monte-Carlo simulation on the SFR values of CG galaxies.  The distribution of median offsets for the isolated (blue solid) and embedded (red dashed) CG galaxies is shown after 10,000 resamplings of the error distribution.  Both control and CG galaxies were re-drawn, and their offsets recalculated.  The median of the distribution of median offsets for isolated CGs is +0.07$\pm$0.03 dex, where the uncertainty is a 1$\sigma$ width of the distribution.  Embedded CG galaxies show median SFRs of  $-$0.06$\pm$0.06 dex.  These distributions indicate that the effect of errors on the SFRs are not likely to change the results of our analysis.  KS tests indicate a 0\% chance that the two distributions were drawn from the same parent sample.} 
\end{figure}

In the Appendix, we include additional figures and tables for illustrative purposes.  Figures \ref{fig:emb_mosaic} and \ref{fig:iso_mosaic} show Sloan images of sample CGs from the embedded and isolated samples respectively.  Tables \ref{tab:embedded_CG} and \ref{tab:isolated_CG} contain information on the embedded and isolated CG galaxies respectively.  Figure \ref{fig:mc_sfr} shows the results of the Monte Carlo simulation conducted in \S\ref{sec:offsets}.


\begin{table*}
\begin{center}
\begin{tabular}{|c|c|c|c|c|c|}
SDSS ObjID & CG ObjID & Redshift & Stellar Mass & Metallicity & SFR\\
\hline
\hline
~ & ~ & $z$ & log(M$_*$) (M$_\odot$) & 12+ log(O/H) & log(SFR) (M$_\odot$ yr$^{-1})$\\
\hline\\
587735348567736612 & SDSSCGA00374.4 & 0.066 & 9.70 & 8.97 & $-$0.23 \\ 
587732470923591912 & SDSSCGA02222.4 & 0.089 & 9.91 & 8.96 & 0.22 \\ 
587742551749427305 & SDSSCGA01310.2 & 0.072 & 10.09 & 9.09 & 0.58 \\ 
587739629557907538 & SDSSCGA00209.3 & 0.063 & 9.78 & 9.00 & 0.51 \\ 
587736525909393590 & SDSSCGA00436.4 & 0.109 & 10.79 & 9.29 & 1.59 \\ 
587742614558212294 & SDSSCGA00213.1 & 0.127 & 10.85 & 9.23 & 1.29 \\ 
587731499184423092 & SDSSCGA02083.4 & 0.050 & 9.52 & 8.96 & $-$0.43 \\ 
588017702397476999 & SDSSCGA01110.2 & 0.067 & 10.02 & 9.04 & 0.34 \\ 
587741600963297361 & SDSSCGA00933.2 & 0.059 & 10.46 & 8.99 & 0.29 \\ 
587729157905514683 & SDSSCGA01012.4 & 0.069 & 9.64 & 9.01 & $-$0.05 \\ 
587736619327553911 & SDSSCGA01186.2 & 0.089 & 10.13 & 9.03 & 0.61 \\ 
587744727687954659 & SDSSCGA01185.4 & 0.092 & 10.07 & 9.20 & 0.31 \\ 
587729160048148639 & SDSSCGA01232.2 & 0.091 & 9.88 & 8.94 & 0.38 \\ 
587742551749427435 & SDSSCGA01310.3 & 0.071 & 9.51 & 8.91 & 0.03 \\ 
587742062159593500 & SDSSCGA00253.2 & 0.077 & 10.15 & 9.17 & 0.72 \\ 
587739648886505610 & SDSSCGA00838.4 & 0.133 & 10.26 & 9.16 & 0.60 \\ 
587741421632356738 & SDSSCGA00979.4 & 0.051 & 9.59 & 8.99 & $-$0.42 \\ 
588017116128870540 & SDSSCGA01825.1 & 0.071 & 10.04 & 8.96 & 0.62 \\ 
588848901523439828 & SDSSCGA01598.4 & 0.082 & 10.44 & 9.17 & 0.33 \\ 
587729388223201436 & SDSSCGA01434.3 & 0.035 & 9.80 & 9.11 & $-$0.34 \\ 
588017116128870551 & SDSSCGA01825.2 & 0.071 & 10.22 & 9.24 & 0.42 \\ 
587731891650494559 & SDSSCGA02027.5 & 0.065 & 9.73 & 9.04 & $-$0.06 \\ 
587732578296529206 & SDSSCGA00435.4 & 0.048 & 9.61 & 9.05 & $-$0.46 \\ 
587734894367539290 & SDSSCGA01858.4 & 0.095 & 10.00 & 9.05 & 0.45 \\ 
587741489294475468 & SDSSCGA01272.3 & 0.042 & 9.17 & 8.71 & $-$0.83 \\ 
587738372745855292 & SDSSCGA01487.3 & 0.096 & 9.98 & 9.00 & 0.47 \\ 
587734894367539289 & SDSSCGA01858.3 & 0.095 & 10.00 & 9.03 & 0.52 \\ 
588017721713623164 & SDSSCGA00494.1 & 0.082 & 10.69 & 9.31 & 0.96 \\ 
587726033854398529 & SDSSCGA01063.3 & 0.078 & 10.48 & 9.23 & 1.03 \\ 
587745403074248994 & SDSSCGA02266.3 & 0.116 & 10.32 & 9.05 & 0.48 \\ 
588017979961901100 & SDSSCGA00904.4 & 0.067 & 9.99 & 9.10 & 0.08 \\ 
587739609162317939 & SDSSCGA02209.3 & 0.078 & 9.72 & 8.97 & 0.17 \\ 
587742060554879300 & SDSSCGA01274.3 & 0.085 & 10.08 & 9.01 & 0.26 \\ 
587736542020042979 & SDSSCGA01833.2 & 0.059 & 10.10 & 9.12 & 0.02 \\ 
587735696987586726 & SDSSCGA01070.2 & 0.083 & 10.65 & 8.86 & $-$0.44 \\ 
587736584977711632 & SDSSCGA02108.1 & 0.063 & 9.87 & 9.13 & 0.13 \\ 
588017721713688616 & SDSSCGA00494.3 & 0.082 & 9.93 & 9.01 & 0.23 \\ 
587742060554944652 & SDSSCGA01274.4 & 0.082 & 10.66 & 9.07 & 0.74 \\ 
587735696987586734 & SDSSCGA01070.4 & 0.082 & 10.38 & 9.31 & 0.58 \\ 
587741600963297386 & SDSSCGA00933.4 & 0.059 & 9.97 & 9.11 & 0.10 \\ 
587731869627252855 & SDSSCGA02011.4 & 0.140 & 10.31 & 9.15 & 0.85 \\ 
587733429770846395 & SDSSCGA01195.2 & 0.073 & 10.70 & 9.29 & 0.87 \\ 
587738065653334220 & SDSSCGA01301.2 & 0.077 & 10.20 & 9.14 & 0.65 \\ 
587741490354389310 & SDSSCGA01109.4 & 0.091 & 10.08 & 9.13 & 0.43 \\ 
588023722317840542 & SDSSCGA00728.4 & 0.044 & 9.30 & 8.91 & $-$0.41 \\ 
588023722317840548 & SDSSCGA00728.3 & 0.044 & 9.90 & 9.07 & $-$0.29 \\ 
587745403074314378 & SDSSCGA02266.4 & 0.116 & 10.04 & 9.10 & 0.22 \\ 
587733429770911869 & SDSSCGA01195.4 & 0.074 & 9.56 & 8.91 & 0.05 \\ 
588016891713880190 & SDSSCGA01137.3 & 0.056 & 9.64 & 9.11 & $-$0.22 \\ 
587735348025819263 & SDSSCGA01108.4 & 0.054 & 9.81 & 9.15 & 0.06 \\ 
\end{tabular}
\end{center}
\caption{Embedded Compact Group galaxy properties.  SDSS Objid is the unique identifier for that galaxy within Sloan, CG Objid is the Galaxy ID from the M09 catalogue where numbers before the decimal indicate the group ID, and the number after the decimal varies between galaxies within one group.  Metallicities calculations are described in \S \ref{sec:metallicity}.  SFR values come from the calculations of \citet{Brinchmann2004}.}
\label{tab:embedded_CG}
\end{table*}

\begin{table*}
\begin{center}
\begin{tabular}{|c|c|c|c|c|c|}
SDSS ObjID & CG ObjID & Redshift & Stellar Mass & Metallicity & SFR\\
\hline
\hline
~ & ~ & $z$ & log(M$_*$) (M$_\odot$) & 12+ log(O/H) & log(SFR) (M$_\odot$ yr$^{-1})$\\
\hline\\
587739844854939883 & SDSSCGA01177.2 & 0.114 & 10.61 & 9.10 & 0.82 \\ 
588023720707424428 & SDSSCGA01088.4 & 0.061 & 10.11 & 8.98 & $-$0.51 \\ 
587736975813836873 & SDSSCGA00878.2 & 0.030 & 9.97 & 9.22 & 0.72 \\ 
587739405707116711 & SDSSCGA01128.1 & 0.054 & 10.48 & 9.04 & 0.41 \\ 
587732471457186108 & SDSSCGA00146.1 & 0.041 & 10.24 & 9.21 & 0.13 \\ 
587725552264479032 & SDSSCGA01533.2 & 0.066 & 10.16 & 9.18 & 0.35 \\ 
587726101490827509 & SDSSCGA00370.4 & 0.085 & 10.25 & 9.23 & 0.57 \\ 
587725980687925438 & SDSSCGA02231.4 & 0.097 & 10.27 & 9.18 & 0.50 \\ 
588017724946972875 & SDSSCGA00942.4 & 0.083 & 9.79 & 8.87 & $-$0.07 \\ 
587735344792928602 & SDSSCGA00422.5 & 0.046 & 9.23 & 8.92 & $-$0.48 \\ 
587732702864080970 & SDSSCGA02097.1 & 0.052 & 9.67 & 8.87 & 0.39 \\ 
587736543623446530 & SDSSCGA00970.4 & 0.087 & 9.89 & 8.97 & 0.57 \\ 
587725041698799708 & SDSSCGA01076.4 & 0.063 & 9.61 & 8.87 & 0.12 \\ 
587725041698799706 & SDSSCGA01076.3 & 0.064 & 9.62 & 8.87 & 0.18 \\ 
587729386610557106 & SDSSCGA01836.3 & 0.072 & 10.08 & 9.09 & $-$0.41 \\ 
587739377773576383 & SDSSCGA00513.3 & 0.079 & 10.13 & 9.19 & 0.49 \\ 
587728930803744828 & SDSSCGA00362.1 & 0.112 & 10.05 & 9.06 & 0.74 \\ 
587745244704473323 & SDSSCGA01087.3 & 0.045 & 9.84 & 9.03 & $-$0.19 \\ 
587742062678900857 & SDSSCGA01309.3 & 0.068 & 9.69 & 8.80 & 0.07 \\ 
587745245240295524 & SDSSCGA00601.3 & 0.113 & 10.36 & 9.20 & 0.48 \\ 
587739827674022085 & SDSSCGA00041.2 & 0.072 & 10.39 & 9.23 & 0.82 \\ 
587729386068705391 & SDSSCGA00078.3 & 0.056 & 9.37 & 8.84 & 0.01 \\ 
587739844310007987 & SDSSCGA01088.3 & 0.061 & 9.93 & 9.02 & 0.06 \\ 
587731886268088671 & SDSSCGA01783.2 & 0.078 & 10.75 & 9.25 & 1.16 \\ 
587745539432776118 & SDSSCGA02092.1 & 0.052 & 9.48 & 8.80 & $-$0.11 \\ 
587742774022045799 & SDSSCGA01953.4 & 0.104 & 9.88 & 8.99 & 0.34 \\ 
587742550686368002 & SDSSCGA00687.2 & 0.087 & 10.18 & 9.19 & 0.34 \\ 
588017627221721269 & SDSSCGA01937.2 & 0.074 & 10.25 & 9.07 & 0.64 \\ 
588848899373072598 & SDSSCGA00713.3 & 0.118 & 10.50 & 9.01 & 0.50 \\ 
588017990162710837 & SDSSCGA00718.2 & 0.139 & 10.84 & 9.32 & 1.11 \\ 
587732483824746731 & SDSSCGA00305.2 & 0.084 & 9.76 & 8.88 & 0.25 \\ 
587742013820305436 & SDSSCGA01570.2 & 0.106 & 9.95 & 8.90 & 0.84 \\ 
587742577521000590 & SDSSCGA01809.3 & 0.093 & 10.33 & 9.19 & 0.48 \\ 
588017627221721251 & SDSSCGA01937.3 & 0.074 & 10.09 & 9.11 & 0.53 \\ 
588007006048485395 & SDSSCGA00936.1 & 0.074 & 10.19 & 9.07 & 0.80 \\ 
587736975813836874 & SDSSCGA00878.3 & 0.031 & 9.71 & 9.16 & $-$0.50 \\ 
587738947198910752 & SDSSCGA01170.2 & 0.043 & 9.71 & 8.94 & $-$0.19 \\ 
587734893827129437 & SDSSCGA00945.1 & 0.060 & 10.56 & 9.16 & 0.79 \\ 
587741720678826052 & SDSSCGA00562.2 & 0.067 & 9.50 & 8.90 & 0.39 \\ 
587726101490827456 & SDSSCGA00370.3 & 0.086 & 10.20 & 9.00 & 0.56 \\ 
587725469057351815 & SDSSCGA01097.1 & 0.132 & 11.09 & 8.89 & 0.87 \\ 
587736782540964307 & SDSSCGA01542.3 & 0.124 & 10.69 & 9.27 & 0.76 \\ 
587739115237015957 & SDSSCGA01970.3 & 0.099 & 9.74 & 8.88 & 0.21 \\ 
587739610230554751 & SDSSCGA00820.2 & 0.073 & 10.20 & 9.09 & 0.37 \\ 
588017627221721162 & SDSSCGA01937.4 & 0.074 & 10.40 & 9.03 & 1.03 \\ 
588848900472111426 & SDSSCGA01473.2 & 0.076 & 10.11 & 9.18 & 0.48 \\ 
587739405707116699 & SDSSCGA01128.2 & 0.054 & 10.46 & 9.08 & 0.67 \\ 
588848899392340185 & SDSSCGA00476.3 & 0.093 & 10.44 & 9.21 & 0.62 \\ 
587729407546228763 & SDSSCGA02056.3 & 0.029 & 9.82 & 9.02 & $-$0.46 \\ 
587739811028992231 & SDSSCGA01347.2 & 0.068 & 9.57 & 8.93 & 0.23 \\ 
587722984441119019 & SDSSCGA01703.2 & 0.120 & 10.36 & 9.13 & 0.74 \\ 
587722984441119010 & SDSSCGA01703.3 & 0.121 & 10.68 & 9.27 & 0.72 \\ 
587736546852864070 & SDSSCGA00337.4 & 0.069 & 9.79 & 9.01 & 0.49 \\ 
587737810644697402 & SDSSCGA01583.3 & 0.103 & 9.81 & 8.95 & 0.46 \\ 
587742578054201471 & SDSSCGA01218.3 & 0.099 & 10.53 & 9.20 & 0.52 \\ 
588016878825701672 & SDSSCGA00227.3 & 0.082 & 10.12 & 9.05 & 0.71 \\ 
587739610230554752 & SDSSCGA00820.3 & 0.072 & 9.91 & 8.92 & 0.35 \\ 
588009368008261825 & SDSSCGA02225.5 & 0.092 & 10.13 & 9.04 & 0.44 \\ 
587738411408294053 & SDSSCGA00614.3 & 0.079 & 10.00 & 8.93 & 0.33 \\ 
587742013297131696 & SDSSCGA01348.3 & 0.102 & 9.96 & 8.96 & 0.54 \\ 
587741603095707807 & SDSSCGA00610.3 & 0.081 & 10.08 & 8.99 & 0.19 \\ 
587742774555508821 & SDSSCGA01132.3 & 0.123 & 10.22 & 9.14 & 0.59 \\ 
\end{tabular}
\caption{\label{tab:isolated_CG}Same as Table \ref{tab:embedded_CG}, but for the isolated CG galaxies}
\end{center}
\end{table*}
\end{appendix}

\end{document}